\documentclass[preprintnumbers,amsmath,amssymb,superscriptaddress]{revtex4}
\usepackage{bm}
\usepackage{graphicx}
\usepackage{dcolumn}
\usepackage{xcolor}
\usepackage{hyperref}
\usepackage{times}
\usepackage{dcolumn}
\usepackage{bm}
\usepackage{dsfont}
\usepackage{mathrsfs}
\usepackage{amsmath}
\usepackage{amsthm}
\usepackage{braket}

\newcommand{\Rmnum}[1]{\expandafter\@slowromancap\romannumeral #1@}

\begin{document}

\title{Quantifying Multipartite Entanglement Based on unified entropy}

\author{Yan Hong}
\affiliation{School of Mathematics and Science, Hebei GEO University, Shijiazhuang 052161, China}

\author{Tongtong Xu}
\affiliation{School of Mathematics and Science, Hebei GEO University, Shijiazhuang 052161, China}

\author{Limin Gao}
\email{gaoliminabc@163.com}
\thanks{Lead contact.}
\affiliation{School of Mathematics and Science, Hebei GEO University, Shijiazhuang 052161, China}
\affiliation{Hebei Province Key Laboratory of Intelligent Sensing and Data Processing for Geo-environment, Hebei GEO University, Shijiazhuang, 052161, China}

\begin{abstract}
In this paper, we investigate the characterization of multipartite entanglement based on the unified $(q,s)$-entropy framework. For bipartite quantum states, we propose an entanglement measure $E_{q,s}^{A|B}(\rho)$ based on unified entropy and derive several analytical lower bounds for it using local orthonormal observables. We demonstrate with concrete examples that our lower bounds provide tighter estimates of quantum entanglement than several existing analytical bounds. For multipartite quantum states, based on unified entropy, we propose two measures $A^k_{q,s}(\rho)$ and $G^k_{q,s}(\rho)$ for quantifying $k$-nonseparability, and prove that these measures satisfy several desirable properties, such as faithfulness, local unitary invariance, and convexity. Moreover, we rigorously prove that when the parameters $q$ and $s$ are in certain ranges, $A^k_{q,s}(\rho)$ and $G^k_{q,s}(\rho)$ also satisfy monotonicity and strong monotonicity. Furthermore, we establish the order relations of the measures $A^k_{q,s}(\rho)$ and $G^k_{q,s}(\rho)$ with respect to the parameters $q$ and $s$ for fixed quantum states. In addition, for fixed $q$ and $s$, we also give their order relations for any two pure states.
\end{abstract}

\maketitle

\section{ Introduction }
Quantum entanglement is not only a distinctive feature that differentiates quantum mechanics from classical physics, but also an essential resource for quantum information processing, such as quantum teleportation \cite{BennettBrassardCrepeauJozsaPeresWoottersPRL1993,YanZhangEPJB2004,GaoYanLiEPL2008}, quantum dense coding \cite{BennettWiesnerPRL1992,MattleWeinfurterKwiatZeilingerPRL1996}, and quantum secret sharing \cite{HilleryBuzekBerthiaumePRA1999}.
In practice, however, the performance of many quantum information processing tasks is closely related to the amount of entanglement available. Consequently, the accurate quantification of quantum entanglement remains a central problem in entanglement theory.

In bipartite quantum systems, substantial progress has been made in the quantification of entanglement \cite{HorodeckiRMP2009,GuhneToth2009,HillWoottersPRL1997,WoottersPRL1998,RungtaBuzekCavesHilleryMilburnPRA2001,
BennettDiVincenzoSmolinWoottersPRA1996,HorodeckiQIC2001,ZyczkowskiHorodeckiSanperaLewensteinPRA1998,VidalWernerPRA2002,KimPRA2010,
GourBandyopadhyaySandersJMP2007,KimSandersJPA2010,YangLuoYangFeiPRA2021,ZhouShenXuanWangFeiAQT2025,XuanShenZhouWangFeiAQT2025,WuTangPRA2026,BrydgesScience2019},  including concurrence \cite{HillWoottersPRL1997,WoottersPRL1998,RungtaBuzekCavesHilleryMilburnPRA2001}, entanglement of formation \cite{BennettDiVincenzoSmolinWoottersPRA1996,HorodeckiQIC2001}, negativity \cite{ZyczkowskiHorodeckiSanperaLewensteinPRA1998,VidalWernerPRA2002}, and entropy-based measures \cite{KimPRA2010, GourBandyopadhyaySandersJMP2007,KimSandersJPA2010,BrydgesScience2019}. For 2-qubit systems, Hill \textit{et al.} introduced the well-known concurrence as an entanglement measure \cite{HillWoottersPRL1997}. A key advantage of this measure is its analytic computability, which allows the entanglement of formation to be explicitly derived from the concurrence \cite{WoottersPRL1998}. However, for most bipartite entanglement measures, the evaluation of entanglement measures for mixed states typically involves nontrivial optimization problems, thereby making it difficult to obtain analytical and computable expressions. Therefore, the search for reliable lower bounds for such measures has become an important research topic. So far, several analytical lower bounds have been proposed based on various theoretical approaches \cite{YangLuoYangFeiPRA2021,WeiLuoFeiQIP2022,BaoZhuQIP2025,WeiFeiJPA2022,ShiSunQIP2023,WangFeiPRA2025}. For example, Yang \textit{et al.} introduced a bipartite entanglement measure $q$-concurrence and, based on the partial transpose and matrix realignment, proposed a lower bound for this measure \cite{YangLuoYangFeiPRA2021}. Subsequently, Wei \textit{et al.} employed similar approaches and gave improved lower bounds, all of which are superior to the original bound obtained by Yang \textit{et al.} \cite{WeiLuoFeiQIP2022}.
Similarly, for another measure $\alpha$-concurrence, Wei \textit{et al.} employed partial transpose and matrix realignment to derive a lower bound \cite{WeiFeiJPA2022}.

In multipartite quantum systems, the intrinsic structural complexity of multipartite quantum states poses a fundamental challenge. To gain deeper insight into the hierarchical structure of multipartite entanglement, researchers have proposed several classification schemes, such as
$k$-nonseparability, $k$-partite entanglement, and $k$-stretchability \cite{HorodeckiRMP2009,GuhneToth2009,Szalay2019Quantum,RenLiSmerziGessnerPRL2021,SzalayToth2025Quantum}. Although these approaches focus on different aspects and can characterize different hierarchical structures of multipartite entanglement, they nevertheless present a formidable challenge to the construction of a unified quantitative framework. Several measures have been proposed to quantify genuine multipartite entanglement, including the GME concurrence, the geometric mean of $q$-concurrence, and the $L$ entropy \cite{MaChenChenSpenglerGabrielHuberPRA2011,ShiChenAP2023,
BasakMalvimatYoonPRL2026}. Inspired by the work of Ma \textit{et al.}, Hong \textit{et al.} proposed $k$-ME concurrence and $(k+1)$-PE concurrence, which respectively characterize the $k$-nonseparability and $(k+1)$-partite entanglement of multipartite quantum states, thereby revealing the hierarchical features of entanglement structures in finer detail \cite{HongGaoYanPRA2012,HongQiGaoYanEPJP2023}. Li \textit{et al.} further introduced $q$-$k$-ME concurrence, $\alpha$-$k$-ME concurrence, $q$-$k$-GM concurrence, $\alpha$-$k$-GM concurrence, $q$-$(k+1)$-PE concurrence, and $\alpha$-$(k+1)$-PE concurrence as measures of $k$-nonseparability or $(k+1)$-partite entanglement, thereby providing a more comprehensive and nuanced description of multipartite entanglement \cite{LiGaoYanPRA2024,LiGaoYanADT2025,LiGaoYanCJP2025}. Beckey \textit{et al.} proposed the concentratable entanglement to quantify the degree of entanglement of multipartite quantum states, and constructed lower bounds for this measure \cite{BeckeyGigenaColesCerezoPRL2021,BeckeyPelegriFouldsPearsonPRA2023}. In Ref. \cite{LuoPRA2025}, Luo \textit{et al.}  generalized concentratable entanglement to unified-entropy concentratable entanglement $E_{\alpha,\beta}^{(s)}(\rho)$ within the unified entropy framework, and employed measure $E_{\alpha,\beta}^{(s)}(\rho)$ as a quantifier of entanglement resources. They further revealed its order relation with respect to the parameter $\alpha$ \cite{LuoPRA2025}.

In this paper, we investigate the characterization of multipartite entanglement based on unified entropy. Inspired by Refs. \cite{MaChenChenSpenglerGabrielHuberPRA2011,HongGaoYanPRA2012,HongQiGaoYanEPJP2023,ShiChenAP2023,LiGaoYanPRA2024,LiGaoYanADT2025,LiGaoYanCJP2025,LuoPRA2025}, we construct two distinct measures of $k$-nonseparability based on unified $(q,s)$-entropy. For bipartite quantum systems, we derive several analytical lower bounds for the proposed entanglement measure, while for multipartite systems, we establish a series of order relations for both measures with respect to the parameters $q$ and $s$, as well as the ordering relations between arbitrary pure states.

This paper is organized as follows. In Sec. II, we present the necessary preliminaries. In Sec. III, based on unified entropy, we define a bipartite entanglement measure, prove that it satisfies strong monotonicity, and derive a series of analytical lower bounds via local orthonormal observables. In Sec. IV, we first construct two distinct measures of $k$-nonseparability using unified entropy and demonstrate that they possess several desirable properties; we then establish a series of order relations for both measures with respect to the parameters $q$ and $s$, as well as for pure states. In Sec. V, we provide a summary.

\section{Preliminaries }

For a quantum state $\rho$ acting on a Hilbert space $H$ with $\dim H=d$, the quantum unified $(q,s)$-entropy is defined as
\begin{equation}
S_{q,s}(\rho)=
\begin{cases}
\dfrac{\big(\operatorname{Tr}\rho^q\big)^s-1}{(1-q)s},
& q>0,\ q\neq1,\ s\neq0,\\[2mm]
\dfrac{\ln\operatorname{Tr}\rho^q}{1-q},
& q>0,\ q\neq1,\ s=0,\\[2mm]
-\operatorname{Tr}(\rho\ln\rho),
& q=1.
\end{cases}
\end{equation}
In the limits $s\rightarrow0$ and $q\rightarrow1$,
the unified $(q,s)$-entropy reduces to the R\'enyi entropy
and the von Neumann entropy, respectively.

The quantum unified $(q,s)$-entropy $S_{q,s}(\rho)$ satisfies the following properties \cite{HuYeJMP2006,RasteginJSP2011,LuoPRA2025}:

(P1) The quantum unified $(q,s)$-entropy satisfies
$S_{q,s}(\rho)\geq0$ for any state $\rho$.
Moreover, $S_{q,s}(\rho)=0$ if and only if $\rho$ is a pure state.

(P2) The quantum unified $(q,s)$-entropy is invariant under unitary transformations, i.e.,
\begin{equation}
S_{q,s}(\rho)=S_{q,s}(U\rho U^\dagger),
\end{equation}
where $U$ is an arbitrary unitary operator.

(P3) Let
\begin{equation}
\begin{split}
\mathcal{C}={}&\{(q,s):0<q\leq1,\ 0<qs\leq1\}\\
&\cup\{(q,s):q\geq1,\ qs\geq1\}\\
&\cup\{(q,s):0<q<1,\ 0\leq s\leq1\}.
\end{split}
\end{equation}
The three parameter regions are obtained from Refs.~\cite{HuYeJMP2006} and \cite{RasteginJSP2011}, respectively.
The quantum unified $(q,s)$-entropy $S_{q,s}(\rho)$ is concave
for any $(q,s)\in \mathcal{C}$, i.e.,
\begin{equation}
S_{q,s}\left(\sum_i p_i\rho_i\right)
\geq
\sum_i p_i S_{q,s}(\rho_i),
\end{equation}
where $p_i\geq0$, $\sum_i p_i=1$, and $\rho_i$ are quantum states.

A collection
$\gamma=\{\gamma_1,\gamma_2,\ldots,\gamma_k\}$
is called a $k$-partition of $\{1,2,\ldots,N\}$ if
\begin{equation}\label{k-partition}
\gamma_i\neq\varnothing,\qquad
\gamma_i\cap\gamma_j=\varnothing\quad(i\neq j),
\qquad
\bigcup_{i=1}^{k}\gamma_i=\{1,2,\ldots,N\}.
\end{equation}
Let $P_k$ be a set consisting of all possible $k$-partitions of  $\{1,2,\cdots,N\}$, i.e. $P_k:=\{\gamma:\gamma \text{ is a } k\text{-partition of }\{1,2,\ldots,N\}\}.$

An $N$-partite  pure state $|\psi\rangle\in H_1\otimes H_2\otimes\cdots\otimes H_N$ is  $k$-separable if there exists a $k$-partition $\{\gamma_1,\gamma_2,\cdots,\gamma_k\}$ such that $|\psi\rangle$ can be expressed as
$|\psi\rangle=|\psi_{\gamma_1}\rangle|\psi_{\gamma_2}\rangle\cdots|\psi_{\gamma_k}\rangle$.
An $N$-partite mixed state $\rho$ is called $k$-separable if it can be written as a convex combination of $k$-separable pure states, that is, $\rho=\sum\limits_lp_l|\psi^{(l)}\rangle\langle\psi^{(l)}|,$ where each $|\psi^{(l)}\rangle$ may be separable with respect to a different $k$-partition.
If a quantum state $\rho$ is not $k$-separable, it is called $k$-nonseparable. In the extreme case $k=N$, the conventional notion of full
separability is recovered: an $N$-separable state is precisely a fully separable state.

\section{Entanglement measure for bipartite states and its lower bounds}

\subsection{Entanglement measure for bipartite quantum states via  quantum unified $(q,s)$-entropy}

For an arbitrary bipartite pure state $|\psi\rangle\in H_A\otimes H_B$, the unified $(q,s)$-entropy-based entanglement measure is defined as
\begin{equation*}\label{Cd1}
\begin{array}{rl}
C_q(|\psi\rangle)=1-\operatorname{Tr}\rho_A^q
\end{array}
\end{equation*}
for $q>1$ \cite{YangLuoYangFeiPRA2021,LiGaoYanPRA2024}, and as
\begin{equation*}\label{Cd2}
\begin{array}{rl}
C_\alpha(|\psi\rangle)=\operatorname{Tr}\rho_A^\alpha-1
\end{array}
\end{equation*}
for $0<\alpha<1$ \cite{WeiFeiJPA2022,LiGaoYanPRA2024}.
For an arbitrary bipartite mixed state $\rho\in H_A\otimes H_B$, the corresponding parameterized entanglement measures
$C_q(\rho)$ and $C_\alpha(\rho)$ are defined via the convex-roof extension,
\begin{equation*}\label{Cd3}
\begin{array}{rl}
C_q(\rho)=\min\limits_{\{p_l,|\psi^{(l)}\rangle\}}\sum\limits_{l}p_lC_q(|\psi^{(l)}\rangle)
\end{array}
\end{equation*}
for $q>1$ \cite{YangLuoYangFeiPRA2021,LiGaoYanPRA2024}, and
\begin{equation*}\label{Cd4}
\begin{array}{rl}
C_\alpha(\rho)=\min\limits_{\{p_l,|\psi^{(l)}\rangle\}}\sum\limits_{l}p_lC_\alpha(|\psi^{(l)}\rangle)
\end{array}
\end{equation*}
for $0<\alpha<1$ \cite{WeiFeiJPA2022,LiGaoYanPRA2024},
where the minimum is taken over all pure-state decompositions $\{p_l,|\psi^{(l)}\rangle\}$ of $\rho$.

These two measures are respectively referred to as the $q$-concurrence and $\alpha$-concurrence.

Motivated by these measures, we next  construct an entanglement measure of bipartite quantum states based on  quantum unified $(q,s)$-entropy.

\emph{Definition 1.}
For any bipartite pure state $|\psi\rangle\in H_A\otimes H_B$, we define the unified $(q,s)$-entropy entanglement as
\begin{equation}\label{D3}
E_{q,s}^{A|B}(|\psi\rangle)
=
S_{q,s}(\rho_A),
\end{equation}
where $\rho_A=\operatorname{Tr}_B(|\psi\rangle\langle\psi|)$.

By convex roof construction, the above measure of pure states is extended to arbitrary mixed states $\rho\in H_A\otimes H_{B}$ with $\dim H_A=d_A$ and $\dim H_B=d_B$, and for a mixed state $\rho$, it is defined as
\begin{equation}\label{D4}
E_{q,s}^{A|B}(\rho)=\min\limits_{\{p_l,|\psi^{(l)}\rangle\}}\sum\limits_{l}p_lE_{q,s}^{A|B}(|\psi^{(l)}\rangle),
\end{equation}
where the minimum is taken over all pure-state decompositions $\{p_l,|\psi^{(l)}\rangle\}$ of $\rho$.

For a bipartite quantum system $H_A\otimes H_{B}$, Vidal \cite{Vidal2000} proved that if an entanglement measure $E$ meets conditions (a), (b1), (b2), and (c), then it satisfies strong monotonicity.

(a) $E(\rho)\geq0$, and $E(\rho)=0$ if and only if $\rho$ is separable.

(b) For any pure state $|\psi\rangle$, the measure is a function of the reduced density operator $\rho_A=\operatorname{Tr}_{B}(|\psi\rangle\langle\psi|)$ as $E(|\psi\rangle)=f(\rho_A)$, and the function $f$ satisfies the following conditions: (b1)
 $f$ is invariant under any unitary transformations, that is,   $f(\rho_A)=f(U\rho_A U^\dagger)$  where $U$  is an arbitrary unitary operator.
(b2) $f$ is concave, that is,   $f[\lambda\rho_1+(1-\lambda)\rho_2]\geq \lambda f(\rho_1)+(1-\lambda)f(\rho_2)$  where $0\leq\lambda\leq1$ and $\rho_1\in H_A, \rho_2 \in H_A$.

(c) For the mixed states $\rho\in H_A\otimes H_{B}$, the  measure  $E(\rho)$ is defined by convex roof construction.

\emph{Theorem 1.}
For any $(q,s)\in\mathcal{C}$, the bipartite entanglement measure $E_{q,s}^{A|B}$ is strongly monotonic under LOCC. More precisely, for any LOCC instrument producing the post-measurement states $\{\sigma^{(l)}\}$ with probabilities $\{p_l\}$,
\begin{equation}
E_{q,s}^{A|B}(\rho)
\geq
\sum_l p_l E_{q,s}^{A|B}(\sigma^{(l)}).
\end{equation}

\emph{Proof.} Using property  (P1) of the quantum unified $(q,s)$-entropy,  Eqs. (\ref{D3}) and (\ref{D4}), we obtain directly that the  measure  $E_{q,s}^{A|B}(\rho)$ satisfies condition (a).
By  property  (P2) of the quantum unified $(q,s)$-entropy, we derive that the  measure  $E_{q,s}^{A|B}(\rho)$ satisfies condition (b1).
The  measure  $E_{q,s}^{A|B}(\rho)$ satisfies condition (b2) by  property  (P3) of the quantum unified $(q,s)$-entropy.
From Eq.  (\ref{D4}), we can  easily see that the  measure  $E_{q,s}^{A|B}(\rho)$ is defined by convex roof construction for the mixed states $\rho\in H_A\otimes H_{B}$. So, the  measure  $E_{q,s}^{A|B}(\rho)$ satisfies conditions (a), (b1), (b2), and (c). Hence the  measure  $E_{q,s}^{A|B}(\rho)$ satisfies strong monotonicity under LOCC operations. When $q=1$, the unified entropy reduces to the von Neumann entropy, and the above conclusion follows from the corresponding properties of the entropy of entanglement.\hfill$\blacksquare$

\subsection{Lower bounds for the entanglement measure $E_{q,s}^{A|B}(\rho)$ via local orthonormal observables}

For a two-qudit state
$\rho\in H_A\otimes H_B$ with
$H_A=H_B=H$ and $\dim H=d$, we derive lower bounds for the
bipartite entanglement measure $E_{q,s}^{A|B}(\rho)$ in terms of
local orthonormal observables.

Let
$\{B_u:u=1,2,\ldots,d^2\}$ be a complete set of local orthonormal
observables on $H$, satisfying
\[
\operatorname{Tr}(B_uB_v)=\delta_{uv}.
\]
Here $\{B_u\}$ forms an orthonormal basis of the real Hilbert space of Hermitian operators equipped with the
Hilbert--Schmidt inner product on $H$. Let
$\{|e_u\rangle:u=1,2,\ldots,d^2\}$ be an orthonormal basis of
$\mathbb{C}^{d^2}$. For a bipartite state $\rho$, define the operator
$L(\rho)$ as
\begin{equation}
L(\rho)
=
\sum_{u,v=1}^{d^2}
\operatorname{Tr}(B_u\otimes B_v\rho)
|e_u\rangle\langle e_v|.
\end{equation}
It immediately follows from the linearity of the trace that
\begin{equation}
L\left(\sum_l p_l\rho_l\right)
=
\sum_l p_lL(\rho_l).
\end{equation}

For a pure state $|\psi\rangle\in H_A\otimes H_B$,
its Schmidt decomposition can be written as
\[
|\psi\rangle
=
\sum_{m=1}^{d}\lambda_m
|f_m\rangle|h_m\rangle,
\]
where $\{\lambda_m\geq0\}$ are the Schmidt coefficients satisfying
$\sum_m\lambda_m^2=1$.

\emph{Proposition 1.}
For any two-qudit pure state
$|\psi\rangle\in H_A\otimes H_B$ with
$H_A=H_B=H$ and $\dim H=d$, one has
\begin{equation}\label{L3-0}
\|L(|\psi\rangle\langle\psi|)\|_{\operatorname{Tr}}
=
\|\rho^{T_A}\|_{\operatorname{Tr}},
\end{equation}
where $\rho^{T_A}$ denotes the partial transpose of
$\rho=|\psi\rangle\langle\psi|$ with respect to subsystem $A$, and
$\|\cdot\|_{\operatorname{Tr}}$ represents the trace norm.

The proof is provided in Appendix A.

On Hilbert space $ H_A\otimes H_B$ with $\dim H_A=d_A$ and $\dim H_B=d_B$, let $d=\min\{d_A,d_B\}$.
For any bipartite mixed state $\rho\in H_A\otimes H_B$, Yang \textit{et al.} proposed a lower bound of the $q$-concurrence
\begin{equation}\label{Cbound1}
\begin{array}{rl}
C_q(\rho)\geq\dfrac{\Big(\max\Big\{\|\rho^{T_A}\|_{\operatorname{Tr}}^{q-1},\|\mathcal{R}(\rho)\|_{\operatorname{Tr}}^{q-1}\Big\}-1\Big)^2}{d^{2q-2}-d^{q-1}}
\end{array}
\end{equation}
for $q\geq2$ \cite{YangLuoYangFeiPRA2021}. Here $\rho^{T_A}$ is the partial transpose with  respect to subsystem $A$, $\mathcal{R}$ is the matrix realignment operation, $\|\cdot\|_{\operatorname{Tr}}$ denotes the trace norm.
From Eq. (\ref{L3-0}) and inequality (\ref{Cbound1}), we derive a lower bound on bipartite entanglement measure $E_{q,s}^{A|B}(\rho)$ in terms of local orthonormal observables.

\emph{Theorem 2.} On Hilbert space $ H_A\otimes H_B$ with $\dim H_A=\dim H_B=d$, for any 2-qudit quantum state $\rho$ with $\|L(\rho)\|_{\operatorname{Tr}}\geq1$, entanglement measure $E_{q,s}^{A|B}(\rho)$ satisfies the lower bound
\begin{equation}\label{Ebound1-0}
\begin{array}{rl}
E_{q,s}^{A|B}(\rho)\geq\dfrac{1-\left[1-\dfrac{1}{d^{2q-2}-d^{q-1}}\Big(\|L(\rho)\|_{\operatorname{Tr}}^{q-1}-1\Big)^2\right]^s}{(q-1)s}
\end{array}
\end{equation}
for $(q,s)\in\mathcal C$ with $q\ge2$ and $0<s\le1$.

The proof of theorem 2 is presented in Appendix B.

In Ref. \cite{WeiLuoFeiQIP2022}, Wei \textit{et al.} derived lower bounds of the $q$-concurrence
\begin{equation}\label{Cbound2}
\begin{array}{rl}
C_q(\rho)\geq\dfrac{1-2^{1-q}}{2-2^{2-r}}\Big(\max\big\{\|\rho^{T_A}\|_{\operatorname{Tr}},\|\mathcal{R}(\rho)\|_{\operatorname{Tr}}\big\}-1\Big)^2,
\end{array}
\end{equation}
for \(d=2\), where \(r=2.4721\) and \(r\le q<3\), and
\begin{equation}\label{Cbound3}
\begin{array}{rl}
C_q(\rho)\geq\dfrac{1-d^{1-q}}{(d-1)^{2}}\Big(\max\big\{\|\rho^{T_A}\|_{\operatorname{Tr}},\|\mathcal{R}(\rho)\|_{\operatorname{Tr}}\big\}-1\Big)^2
\end{array}
\end{equation}
for either $d=2$ with $q\geq3$ or $d\geq3$ with $q\geq2$.
Combining Eq. (\ref{L3-0}) with inequality (\ref{Cbound2}) and inequality (\ref{Cbound3}), respectively, yields two lower bounds on bipartite entanglement measure $E_{q,s}^{A|B}(\rho)$, as follows:

\emph{Theorem 3.} On Hilbert space $ H_A\otimes H_B$ with $\dim H_A=\dim H_B=d$, for any 2-qudit quantum state $\rho$ with $\|L(\rho)\|_{\operatorname{Tr}}\geq1$, the following lower bounds on entanglement measure $E_{q,s}^{A|B}(\rho)$ can be obtained,
\begin{equation}\label{Ebound2-0}
\begin{array}{rl}
E_{q,s}^{A|B}(\rho)\geq\dfrac{1-\left[1-\dfrac{1-2^{1-q}}{2-2^{2-r}}\Big(\|L(\rho)\|_{\operatorname{Tr}}-1\Big)^2\right]^s}{(q-1)s}
\end{array}
\end{equation}
for $d=2$, $2.4721=r\leq q<3$ and $0<s\leq1$, and
\begin{equation}\label{Ebound3-0}
\begin{array}{rl}
E_{q,s}^{A|B}(\rho)\geq\dfrac{1-\left[1-\dfrac{1-d^{1-q}}{(d-1)^{2}}\Big(\|L(\rho)\|_{\operatorname{Tr}}-1\Big)^2\right]^s}{(q-1)s}
\end{array}
\end{equation}
for $0<s\leq1$, either $d=2$ with $q\geq3$  or $d\geq3$ with $q\geq2$.

The proof of theorem 3 is presented in Appendix B.

The lower bound of the $\alpha$-concurrence is constrained by
\begin{equation}\label{Cbound5}
\begin{array}{rl}
C_\alpha(\rho)\geq\dfrac{d^{1-\alpha}-1}{d-1}\big(\max\big\{\|\rho^{T_A}\|_{\operatorname{Tr}},\|\mathcal{R}(\rho)\|_{\operatorname{Tr}}\big\}-1\big)
\end{array}
\end{equation}
for $0<\alpha\leq\frac{1}{2}$ \cite{WeiFeiJPA2022}.
Applying Eq. (\ref{L3-0}) together with (\ref{Cbound5}), we obtain a lower bound on bipartite entanglement measure $E_{q,s}^{A|B}(\rho)$.

\emph{Theorem 4.}  On Hilbert space $ H_A\otimes H_B$ with $\dim H_A=\dim H_B=d$, for any 2-qudit quantum state $\rho$ with $\|L(\rho)\|_{\operatorname{Tr}}\geq1$,  one has
\begin{equation}\label{Ebound5-0}
\begin{array}{rl}
E_{q,s}^{A|B}(\rho)\geq\dfrac{\left[\dfrac{d^{1-q}-1}{d-1}\big(\|L(\rho)\|_{\operatorname{Tr}}-1\big)+1\right]^{s}-1}{(1-q)s}
\end{array}
\end{equation}
for $0<q\leq\frac{1}{2}$ with $s\geq1$.

The proof of theorem 4 is presented in Appendix B.

\emph{Example 1.} Consider a quantum state $\rho=\dfrac{1}{4}(I-\sum\limits_{i=1}^5|\psi_i\rangle\langle\psi_i|)$ with $|\psi_1\rangle=\dfrac{|0\rangle(|0\rangle-|1\rangle)}{\sqrt{2}},|\psi_2\rangle=\dfrac{(|0\rangle-|1\rangle)|2\rangle}{\sqrt{2}},
|\psi_3\rangle=\dfrac{|2\rangle(|1\rangle-|2\rangle)}{\sqrt{2}},|\psi_4\rangle=\dfrac{(|1\rangle-|2\rangle)|0\rangle}{\sqrt{2}},
|\psi_5\rangle=\dfrac{(|0\rangle+|1\rangle+|2\rangle)^{\otimes2}}{3}$.

For \(s=1\), the unified entropy reduces to $E_{q,1}^{A|B}(\rho)=\frac{\operatorname{Tr}(\rho_A^q)-1}{1-q},$ which implies that $C_q(\rho)=(q-1)E_{q,1}^{A|B}(\rho).$ For \(s=1\) and \(q=2.1\), Theorems 2 and 3 imply, respectively, that $C_q(\rho)=1.1E_{2.1,1}^{A|B}(\rho)\geq 0.00119$ and $C_q(\rho)=1.1E_{2.1,1}^{A|B}(\rho)\geq 0.00134.$

For $s=1$ and $\alpha=0.5$, Theorem~4 gives $C_{0.5}(\rho)=\frac{1}{2}E_{0.5,1}^{A|B}(\rho)\geq0.03199.$
But theorem 1 of Ref. \cite{YangLuoYangFeiPRA2021} and theorem 1 of Ref. \cite{WeiLuoFeiQIP2022} both yield  $C_q(\rho)\geq0$,
and theorem 2 of Ref. \cite{WeiFeiJPA2022} gives $C_\alpha(\rho)\geq0$.
Therefore, for the quantum state $\rho=\dfrac{1}{4}(I-\sum\limits_{i=1}^5|\psi_i\rangle\langle\psi_i|)$, our theorems 2 and 3 each outperform theorem 1 in Ref. \cite{YangLuoYangFeiPRA2021} and theorem 1 in Ref. \cite{WeiLuoFeiQIP2022}, while theorem 4 outperforms Theorem 2 in Ref. \cite{WeiFeiJPA2022}.

\section{ Two measures of $k$-nonseparability for multipartite states and ordering relations }

\subsection{Two measures of $k$-nonseparability for multipartite states via  quantum unified $(q,s)$-entropy}

In this section, we will construct two measures for quantifying $k$-nonseparability based on  quantum unified $(q,s)$-entropy.

\emph{Definition 2.} For any $N$-partite pure state $|\psi\rangle\in H_1\otimes H_2\otimes\cdots\otimes H_N$ with $\dim H_i=d_i$,  the arithmetic mean measure of $k$-nonseparability based on quantum unified $(q,s)$-entropy is defined as
\begin{equation}\label{D1}
A^k_{q,s}(|\psi\rangle)=\min\limits_{\gamma\in P_k}\dfrac{\sum\limits_{t=1}^{k}S_{q,s}(\rho_{\gamma_t})}{k},
\end{equation}
and the geometric  mean measure of $k$-nonseparability based on quantum unified $(q,s)$-entropy is defined as
\begin{equation}\label{D1-1}
G^k_{q,s}(|\psi\rangle)=\Big(\prod\limits_{\gamma\in P_k}\big[\sum\limits_{t=1}^{k}S_{q,s}(\rho_{\gamma_t})/k\big]\Big)^{\frac{1}{c(k)}}.
\end{equation}
Here $\rho_{\gamma_t}$ is the reduced density operator of pure state $|\psi\rangle$ for subsystem $\gamma_t$, both the minimum and the product are taken over all possible $k$-partitions. Here $c(k)=S(N,k)$ denotes the Stirling number of the second kind, i.e., the number of $k$-partitions of $\{1,2,\ldots,N\}$.

By convex roof construction, the above  measures of $k$-nonseparability for pure states are extended to arbitrary mixed states $\rho$.  For a mixed state $\rho$,  we define the arithmetic mean $k$-nonseparability measure based on  quantum unified $(q,s)$-entropy  as
\begin{equation}\label{D2}
A^k_{q,s}(\rho)=\min\limits_{\{p_l,|\psi^{(l)}\rangle\}}\sum\limits_{l}p_lA^k_{q,s}(|\psi^{(l)}\rangle),
\end{equation}
and the geometric  mean measure of $k$-nonseparability based on quantum unified $(q,s)$-entropy  as
\begin{equation}\label{D2-1}
G^k_{q,s}(\rho)=\min\limits_{\{p_l,|\psi^{(l)}\rangle\}}\sum\limits_{l}p_lG^k_{q,s}(|\psi^{(l)}\rangle),
\end{equation}
where the minimum is taken over all pure-state decompositions $\{p_l,|\psi^{(l)}\rangle\}$ of $\rho$.

\emph{Theorem 5.} Let $\rho$ be  arbitrary quantum state in $N$-partite quantum systems $H_1\otimes H_2\otimes\cdots\otimes H_N$ with $\dim H_i=d_i$. Let $A^k_{q,s}(\rho)$ and $G^k_{q,s}(\rho)$ denote arithmetic mean measure and the geometric-mean measure of $k$-nonseparability, respectively, which have the following properties:

(i) (Faithfulness) (ia) $A^k_{q,s}(\rho)\geq0$.    $A^k_{q,s}(\rho)=0$ if and only if $\rho$ is $k$-separable.
(ib) $G^k_{q,s}(\rho)\geq0$.    $G^k_{q,s}(\rho)=0$ if and only if $\rho$ is $k$-separable.

(ii) (Invariant under local unitary transformations) Both $A^k_{q,s}(\rho)$ and $G^k_{q,s}(\rho)$  are invariants under local unitary transformations, namely, $A^k_{q,s}(\rho)=A^k_{q,s}((U_1\otimes\cdots\otimes U_N)
\rho
(U_1^\dagger\otimes\cdots\otimes U_N^\dagger))$ and $G^k_{q,s}(\rho)=G^k_{q,s}((U_1\otimes\cdots\otimes U_N)
\rho
(U_1^\dagger\otimes\cdots\otimes U_N^\dagger)).$
where $U_i$  is an arbitrary unitary operator acting on subsystem $H_i$.

(iii) (Convexity) Both $A^k_{q,s}(\rho)$ and $G^k_{q,s}(\rho)$  are convex,   $A^k_{q,s}(\sum\limits_ip_i\rho_i)\leq \sum\limits_ip_iA^k_{q,s}(\rho_i)$  and  $G^k_{q,s}(\sum\limits_ip_i\rho_i)\leq \sum\limits_ip_iG^k_{q,s}(\rho_i)$,
where $p_i\geq0$ and $\sum\limits_ip_i=1.$

(iv) (Monotonicity) For any $(q,s)\in \mathcal{C}$, both $A^k_{q,s}(\rho)$  and $G^k_{q,s}(\rho)$ are  nonincreasing under local operation and classical communication (LOCC) $\Lambda_{\rm LOCC}$, i.e.,
$A^k_{q,s}(\rho)\geq A^k_{q,s}[\Lambda_{\rm LOCC}(\rho)]$ and $G^k_{q,s}(\rho)\geq G^k_{q,s}[\Lambda_{\rm LOCC}(\rho)]$
for any $(q,s)\in \mathcal{C}$.

(v) (Strong monotonicity) For any $(q,s)\in \mathcal{C}$,  both $A^k_{q,s}(\rho)$  and $G^k_{q,s}(\rho)$ are average nonincreasing under LOCC, i.e.,
$A^k_{q,s}(\rho)\geq \sum\limits_ip_iA^k_{q,s}(\sigma_i)$  and $G^k_{q,s}(\rho)\geq \sum\limits_ip_iG^k_{q,s}(\sigma_i)$ for any $(q,s)\in \mathcal{C}$, where $\sigma_i$ is obtained with probability $p_i$ by applying LOCC $\Lambda_{\rm LOCC}$ to $\rho$.

\emph{Proof:}
(i) We first prove the result for pure states. According to property (P1) of the
unified $(q,s)$-entropy, for any reduced density matrix $\rho_{\gamma_t}$,
\[
S_{q,s}(\rho_{\gamma_t})=0
\]
if and only if $\rho_{\gamma_t}$ is pure. Hence, for a pure state
$|\psi\rangle$,
\[
\sum_{t=1}^{k}S_{q,s}(\rho_{\gamma_t})=0
\]
holds if and only if $|\psi\rangle$ is a product state with respect
to the corresponding $k$-partition
$\gamma=\{\gamma_1,\ldots,\gamma_k\}$. Therefore, both definitions
of $A^k_{q,s}$ and $G^k_{q,s}$ give
\[
A^k_{q,s}(|\psi\rangle)=0
\quad\Longleftrightarrow\quad
G^k_{q,s}(|\psi\rangle)=0
\quad\Longleftrightarrow\quad
|\psi\rangle\ \text{is $k$-separable}.
\]
Indeed, for $A^k_{q,s}$ the result follows directly from the
minimization over all $k$-partitions.
For pure states, $G^k_{q,s}(|\psi\rangle)=0$ implies that at least one
$k$-partition gives a vanishing arithmetic mean. Therefore,
$|\psi\rangle$ is separable with respect to this partition and hence
$k$-separable.

The result can then be extended to mixed states. The set of $k$-separable mixed states is defined as the convex hull
of all k-separable pure states with arbitrary partitions. Let
\[
E^k(\rho)\in\{A^k_{q,s}(\rho),G^k_{q,s}(\rho)\}.
\]

Let $\{p_l,|\psi^{(l)}\rangle\}$ be an optimal decomposition attaining the
convex roof.
\[
E^k(\rho)=
\sum_l p_lE^k(|\psi^{(l)}\rangle)
\]
for a pure-state decomposition
\[
\rho=\sum_l p_l|\psi^{(l)}\rangle\langle\psi^{(l)}|.
\]
If
\[
E^k(\rho)=0,
\]
then
\[
\sum_l p_lE^k(|\psi^{(l)}\rangle)=0 .
\]
Since every term is nonnegative,
\[
E^k(|\psi^{(l)}\rangle)=0
\]
for all $l$ with $p_l>0$. From the pure-state result, all
$|\psi^{(l)}\rangle$ are $k$-separable. Therefore, due to the
convexity of the set of $k$-separable states, $\rho$ is
$k$-separable.

Conversely, if $\rho$ is $k$-separable, it can be decomposed into
$k$-separable pure states. Since each pure component satisfies
\[
E^k(|\psi^{(l)}\rangle)=0,
\]
the convex-roof definition immediately gives
\[
E^k(\rho)=0 .
\]
Therefore,
\[
E^k(\rho)=0
\]
if and only if $\rho$ is $k$-separable. Consequently, both
$A^k_{q,s}(\rho)$ and $G^k_{q,s}(\rho)$ are faithful.

(ii) By  property (P2) of quantum unified $(q,s)$-entropy, both $A^k_{q,s}(\rho)$ and $G^k_{q,s}(\rho)$ remain invariant under local unitary transformations.

(iii) Since both $A^k_{q,s}(\rho)$ and $G^k_{q,s}(\rho)$ are defined using the convex roof construction, they are convex.

(iv) First, we prove $A^k_{q,s}(\rho)$ satisfies item (iv).

For any pure state $|\psi\rangle$, we will consider two cases: one in which $\Lambda_{\rm LOCC}(|\psi\rangle\langle\psi|)$ is a pure state, and one in which $\Lambda_{\rm LOCC}(|\psi\rangle\langle\psi|)$ is a mixed state.

Case 1. If $\Lambda_{\rm LOCC}(|\psi\rangle\langle\psi|)$ is a pure state, we have
\begin{equation*}
\begin{array}{rl}
A^k_{q,s}[\Lambda_{\rm LOCC}(|\psi\rangle\langle\psi|)]=&\min\limits_{\gamma\in P_k}\dfrac{\sum\limits_{t=1}^{k}S_{q,s}(\widetilde{\rho}_{\gamma_t})}{k}\\
\leq&\min\limits_{\gamma\in P_k}\dfrac{\sum\limits_{t=1}^{k}S_{q,s}(\rho_{\gamma_t})}{k}\\
=&A^k_{q,s}(|\psi\rangle).
\end{array}
\end{equation*}
Here $\rho=|\psi\rangle\langle\psi|, \rho_{\gamma_t}=\operatorname{Tr}_{\bar{\gamma}_t}(\rho)$,  $\widetilde{\rho}=|\widetilde{\psi}\rangle\langle\widetilde{\psi}|$ and $\widetilde{\rho}_{\gamma_t}=\operatorname{Tr}_{\bar{\gamma}_t}(\widetilde{\rho})$ with $|\widetilde{\psi}\rangle=\Lambda_{\rm LOCC}(|\psi\rangle\langle\psi|)$.
Both the two equations hold by the definition (\ref{D1}) of the arithmetic mean measure.
The inequality  is based on
\begin{equation}\label{iva4}
S_{q,s}(\widetilde{\rho}_{\gamma_t})\leq S_{q,s}(\rho_{\gamma_t}).
\end{equation}
Inequality (\ref{iva4}) is proved in the Appendix C.

Case 2. If $\Lambda_{\rm LOCC}(|\psi\rangle\langle\psi|)$ is a mixed state, consider an LOCC instrument described by Kraus operators
$\{M_l\}$ with $\sum\limits_lM_l^\dagger M_l=I. $ Then $\Lambda_{\rm LOCC}(|\psi\rangle\langle\psi|)=\sum\limits_lM_l|\psi\rangle\langle\psi|M_l^\dagger=\sum\limits_lp_l|\omega^{(l)}\rangle\langle\omega^{(l)}|$ with $|\omega^{(l)}\rangle=\dfrac{M_l|\psi\rangle}{\sqrt{p_l}}$ and $p_l=\operatorname{Tr}(M_l|\psi\rangle\langle\psi|M_l^\dagger)$.
By regarding $\gamma_t|\bar{\gamma}_t$ as a bipartition, Theorem 1 applies with the subsystem Hilbert space defined as
$H_A=\bigotimes_{i\in\gamma_t}H_i.$ Since $E_{q,s}^{\gamma_t|\bar{\gamma}_t}(|\psi\rangle)=S_{q,s}(\rho_{\gamma_t}),$
where $\bar{\gamma}_t=\{1,2,\cdots,N\}\setminus\gamma_t$, we obtain
\begin{equation*}
\begin{array}{rl}
A^k_{q,s}[\Lambda_{\rm LOCC}(|\psi\rangle\langle\psi|)]\leq&\sum\limits_lp_lA^k_{q,s}(|\omega^{(l)}\rangle)\\
=&\sum\limits_lp_l\min\limits_{\gamma\in P_k}\dfrac{\sum\limits_{t=1}^{k}E_{q,s}^{\gamma_t|\bar{\gamma}_t}(|\omega^{(l)}\rangle)}{k}\\
\leq&\min\limits_{\gamma\in P_k}\sum\limits_lp_l\dfrac{\sum_{t=1}^{k}E_{q,s}^{\gamma_t|\bar{\gamma}_t}(|\omega^{(l)}\rangle)}{k}\\
\leq&\min\limits_{\gamma\in P_k}\dfrac{\sum\limits_{t=1}^{k}E_{q,s}^{\gamma_t|\bar{\gamma}_t}(|\psi\rangle)}{k}\\
=&A^k_{q,s}(|\psi\rangle).
\end{array}
\end{equation*}
The first inequality holds by Eq. (\ref{D2}).
Both the two equations follow from  Eq. (\ref{D1}). The second inequality  holds because
\begin{equation*}
\begin{array}{rl}
&p_1\min\{x_{11},x_{12},\cdots,x_{1n}\}+p_2\min\{x_{21},x_{22},\cdots,x_{2n}\}+\cdots+p_m\min\{x_{m1},x_{m2},\cdots,x_{mn}\}\\
\leq&\min\{p_1x_{11}+p_2x_{21}+\cdots+p_mx_{m1}, p_1x_{12}+p_2x_{22}+\cdots+p_mx_{m2},\cdots,p_1x_{1n}+p_2x_{2n}+\cdots+p_mx_{mn}\}.
\end{array}
\end{equation*}
The third inequality  is based on the inequality
\begin{equation}\label{iva0}
\sum\limits_lp_lE_{q,s}^{\gamma_t|\bar{\gamma}_t}(|\omega^{(l)}\rangle)
\leq E_{q,s}^{\gamma_t|\bar{\gamma}_t}(|\psi\rangle).
\end{equation}
Inequality (\ref{iva0}) is proved in the Appendix D.

Therefore, for any pure state $|\psi\rangle$, we obtain
\begin{equation}\label{ivapure}
A^k_{q,s}[\Lambda_{\rm LOCC}(|\psi\rangle\langle\psi|)]\leq A^k_{q,s}(|\psi\rangle).
\end{equation}

Next, we prove $A^k_{q,s}(\rho)$ satisfies item (iv) for any mixed state $\rho$.
Let $\{p_l,|\psi^{(l)}\rangle\}$ and $\{p_{il},|\psi^{(il)}\rangle\}$  be the optimal decompositions of $A^k_{q,s}(\rho)$ and $A^k_{q,s}[\Lambda_{\rm LOCC}(|\psi^{(l)}\rangle)]$, respectively, in Eq. (\ref{D2}).
Thus, we derive
\begin{equation*}
\begin{array}{rl}
A^k_{q,s}(\rho)=&\sum\limits_lp_lA^k_{q,s}(|\psi^{(l)}\rangle)\\
\geq&\sum\limits_lp_lA^k_{q,s}[\Lambda(|\psi^{(l)}\rangle)]\\
=&\sum\limits_lp_l\sum\limits_ip_{il}A^k_{q,s}(|\psi^{(il)}\rangle)\\
\geq&A^k_{q,s}[\Lambda_{\rm LOCC}(\rho)].
\end{array}
\end{equation*}
Since $\{p_l,|\psi^{(l)}\rangle\}$ and $\{p_{il},|\psi^{(il)}\rangle\}$ are the optimal decompositions of $A^k_{q,s}(\rho)$ and $A^k_{q,s}[\Lambda_{\rm LOCC}(|\psi^{(l)}\rangle)]$, respectively, in Eq. (\ref{D2}), both the two equations hold.
The first inequality  is true by inequality (\ref{ivapure}).
Using the fact that $\{p_lp_{il},|\psi^{(il)}\rangle\}$ is a pure state ensemble decomposition of $\Lambda_{\rm LOCC}(\rho)$, together with Eq. (\ref{D2}), we derive the second inequality.
Hence, for any mixed state $\rho$, we obtain
\begin{equation*}
A^k_{q,s}[\Lambda_{\rm LOCC}(\rho)]\leq A^k_{q,s}(\rho).
\end{equation*}

The proof that $G^k_{q,s}(\rho)$ also satisfies item (iv) is placed in Appendix E.

(v) We first prove that item (v) holds for any pure state $|\psi\rangle$. consider an LOCC instrument described by Kraus operators
$\{M_l\}$ with $\sum\limits_lM_l^\dagger M_l=I$. Then, after applying $\Lambda_{\rm LOCC}$, the pure state $|\psi\rangle$ transforms into $|\omega^{(l)}\rangle$ with probability $p_l$ where $p_l=\operatorname{Tr}(M_l|\psi\rangle\langle\psi|M_l^\dagger)$ and $|\omega^{(l)}\rangle=\dfrac{M_l|\psi\rangle}{\sqrt{p_l}}$.
Suppose that $\{\gamma_1,\gamma_2,\cdots,\gamma_k\}$ is the $k$-partition  such that
$A^k_{q,s}(|\psi\rangle)=\dfrac{\sum\limits_{t=1}^{k}S_{q,s}(\rho_{\gamma_t})}{k},$
 then one has
\begin{equation*}
\begin{array}{rl}
A^k_{q,s}(|\psi\rangle)=&\dfrac{\sum\limits_{t=1}^{k}S_{q,s}(\rho_{\gamma_t})}{k}\\
=&\dfrac{\sum\limits_{t=1}^{k}E_{q,s}^{\gamma_t|\bar{\gamma}_t}(|\psi\rangle)}{k}\\
\geq&\dfrac{\sum\limits_{t=1}^{k}\sum\limits_lp_lE_{q,s}^{\gamma_t|\bar{\gamma}_t}(|\omega^{(l)}\rangle)}{k}\\
=&\sum\limits_lp_l\dfrac{\sum\limits_{t=1}^{k}S_{q,s}(\sigma^{(l)}_{\gamma_t})}{k}\\
\geq&\sum\limits_lp_lA^k_{q,s}(|\omega^{(l)}\rangle)\\
=&\sum\limits_lp_lA^k_{q,s}(\sigma^{(l)}).
\end{array}
\end{equation*}
Here $\rho=|\psi\rangle\langle\psi|, \rho_{\gamma_t}=\operatorname{Tr}_{\bar{\gamma}_t}(\rho)$, $\sigma^{(l)}=|\omega^{(l)}\rangle\langle\omega^{(l)}|$ and
$\sigma^{(l)}_{\gamma_t}=\operatorname{Tr}_{\bar{\gamma}_t}(\sigma^{(l)})$.
The second equation   and third equation both hold by Eq. (\ref{D3}). The first inequality  is valid because of inequality (\ref{iva0}).
The second inequality  is true by Eq. (\ref{D1}).

Next, we prove that item (v) holds for an arbitrary mixed state $\rho$. Let $\{q_i,|\psi^{(i)}\rangle\}$ be the pure  state ensemble decomposition of $\rho$ in  (\ref{D2}) that attains the minimum. consider an LOCC instrument described by Kraus operators $\{M_l\}$ with $\sum\limits_lM_l^\dagger M_l=I$. Then, after applying $\Lambda_{\rm LOCC}$, the mixed state $\rho$ transforms into $\sigma^{(l)}$ with probability $p_{l}$
where $p_{l}=\operatorname{Tr}(M_l\rho M_l^\dagger)$ and $\sigma^{(l)}=\dfrac{M_l\rho M_l^\dagger}{p_{l}}$.
After applying $\Lambda_{\rm LOCC}$, the pure state $|\psi^{(i)}\rangle$ transforms into $|\omega^{(il)}\rangle$ with probability $p_{il}$ where $p_{il}=\operatorname{Tr}(M_l|\psi^{(i)}\rangle\langle\psi^{(i)}|M_l^\dagger)$ and $|\omega^{(il)}\rangle=\dfrac{M_l|\psi^{(i)}\rangle}{\sqrt{p_{il}}}$.
So we have
\begin{equation*}
\begin{array}{rl}
A^k_{q,s}(\rho)=&\sum\limits_iq_iA^k_{q,s}(|\psi^{(i)}\rangle)\\
\geq&\sum\limits_{i,l}q_ip_{il}A^k_{q,s}(|\omega^{(il)}\rangle)\\
=&\sum\limits_lp_l\sum\limits_{i}\frac{q_ip_{il}}{p_l}A^k_{q,s}(|\omega^{(il)}\rangle)\\
\geq&\sum\limits_lp_lA^k_{q,s}(\sigma^{(l)}).
\end{array}
\end{equation*}
For any pure state $|\psi\rangle$, when $(q,s)\in \mathcal{C}$, $A^k_{q,s}(|\psi\rangle)$ does not increase on average under LOCC; therefore, the first inequality holds. The second inequality  holds because  Eq. (\ref{D2}) and the fact that $\{\dfrac{q_ip_{il}}{p_l},|\omega^{(il)}\rangle\}$ is a pure  state ensemble decomposition of $\sigma^{(l)}$.

The proof that $G^k_{q,s}(\rho)$ also satisfies item (v) is placed in Appendix F. \hfill$\blacksquare$

For any $N$-qubit quantum state $\rho\in H_1\otimes H_2\otimes\cdots\otimes H_N$, its permutation invariant part $\rho^{\textrm{PI}}$ is defined as
\begin{equation*}
\begin{array}{rl}
\rho^{\textrm{PI}}= \dfrac{1}{N!}\sum\limits_{i=1}^{N!}\Pi_i\rho\Pi_i^\dagger.
\end{array}
\end{equation*}
Here the sum runs over all permutations $\{\Pi_i\}$ of $N$ particles.
By introducing a permutation invariant part $\rho^{\textrm{PI}}$ of quantum states $\rho$, Gao \textit{et al.} proved that a valid entanglement measure
$E(\rho)$ satisfies $E(\rho)\geq \max\limits_U E(\rho^{\textrm{PI}}_U)$ with the maximum being taken all local unitary transformations and $\rho^{\textrm{PI}}_U=(U\rho U^\dagger)^{\textrm{PI}}$ \cite{GaoPRL2014}. Moreover, Qi \textit{et al.} showed that numerous entanglement measures, such as entanglement of formation, concurrence, geometric-mean measure of entanglement, all satisfy $E(\rho)\geq \max\limits_U E(\rho^{\textrm{PI}}_U)$ \cite{QiGaoYanQIP2021}. It is essential to note that both $k$-nonseparability measures $A^k_{q,s}(\rho)$ and $G^k_{q,s}(\rho)$ we propose satisfy this requirement.

\emph{Theorem 6.}
For any $N$-qubit state $\rho$, the two $k$-nonseparability measures $A^k_{q,s}(\rho)$ and $G^k_{q,s}(\rho)$ satisfy
\begin{equation}\label{AU}
\begin{array}{rl}
A^k_{q,s}(\rho)\geq \max\limits_UA^k_{q,s}(\rho^{\textrm{PI}}_U),
\end{array}
\end{equation}
\begin{equation}\label{GU}
\begin{array}{rl}
G^k_{q,s}(\rho)\geq \max\limits_UG^k_{q,s}(\rho^{\textrm{PI}}_U),
\end{array}
\end{equation}
where the maximum is over all local unitary transformations.

\emph{Proof.} If $\gamma=\{\gamma_1,\gamma_2,\cdots,\gamma_k\}$ is  a $k$-partition of the set $\{1,2\cdots,N\}$, define $\Pi_i(\gamma)=\{\Pi_i(\gamma_1),\cdots,\Pi_i(\gamma_k)\}$ denote the partition obtained from $\gamma$
under the permutation $\Pi_i$. Then $\{\Pi_i(\gamma_1),\Pi_i(\gamma_2),\cdots,\Pi_i(\gamma_k)\}$ is also a $k$-partition of the set $\{1,2\cdots,N\}$.

By the definition of $A^k_{q,s}(|\psi\rangle)$ for pure states, we can obtain
\begin{equation}\label{PIpure}
\begin{array}{rl}
A^k_{q,s}(|\psi\rangle)= A^k_{q,s}(\Pi_i|\psi\rangle).
\end{array}
\end{equation}
Let $\rho^{\mathrm{PI}}$ denote the permutation invariant part
of the pure state density operator $\rho=|\psi\rangle\langle\psi|$. By the convexity of the measure $A^k_{q,s}(\rho)$ and Eq. (\ref{PIpure}), one has
\begin{equation}\label{PIPpure}
\begin{array}{rl}
A^k_{q,s}(\rho^{\textrm{PI}})\leq\dfrac{1}{N!}\sum\limits_{i=1}^{N!}A^k_{q,s}(\Pi_i|\psi\rangle)= A^k_{q,s}(|\psi\rangle).
\end{array}
\end{equation}

For the mixed state $\rho$, suppose that $\{p_l,|\psi^{(l)}\rangle\}$ is the optimal pure state decomposition in the definition of $A^k_{q,s}(\rho)$, it follows from inequality (\ref{PIPpure}) and the convexity of the measure $A^k_{q,s}(\rho)$ that
\begin{equation*}
\begin{array}{rl}
A^k_{q,s}(\rho)=\sum\limits_lp_lA^k_{q,s}(|\psi^{(l)}\rangle)\geq\sum\limits_lp_lA^k_{q,s}(\rho_l^{\textrm{PI}})\geq A^k_{q,s}(\rho^{\textrm{PI}}),
\end{array}
\end{equation*}
where $\rho_l=|\psi^{(l)}\rangle\langle\psi^{(l)}|$. Hence, we have
\begin{equation}\label{PIPmixed}
\begin{array}{rl}
A^k_{q,s}(U\rho U^\dagger)\geq A^k_{q,s}\big((U\rho U^\dagger)^{\textrm{PI}}\big),
\end{array}
\end{equation}
where $U=U_1\otimes U_2\otimes\cdots\otimes U_N$ is an arbitrary local unitary operator.

Utilizing the local unitary invariance of $A^k_{q,s}(\rho)$ and inequality (\ref{PIPmixed}), we obtain
\begin{equation*}
\begin{array}{rl}
A^k_{q,s}(\rho)=A^k_{q,s}(U\rho U^\dagger)\geq A^k_{q,s}\big((U\rho U^\dagger)^{\textrm{PI}}\big).
\end{array}
\end{equation*}
Therefore, one has
\begin{equation*}
\begin{array}{rl}
A^k_{q,s}(\rho)\geq \max\limits_UA^k_{q,s}\big((U\rho U^\dagger)^{\textrm{PI}}\big)=\max\limits_UA^k_{q,s}(\rho_U^{\textrm{PI}}),
\end{array}
\end{equation*}
with $U$ being any local unitary transformation.  Inequality (\ref{GU}) follows by the same argument. \hfill$\blacksquare$

\subsection{The ordering relations of $k$-nonseparability measures $A^k_{q,s}(\rho)$ and $G^k_{q,s}(\rho)$  }

The two $k$-nonseparability measures $A^k_{q,s}(\rho)$ and $G^k_{q,s}(\rho)$ depend on parameters $q$ and $s$. In what follows, we first discuss their order relations with respect to each parameter.

\emph{Theorem 7.} For any $N$-partite quantum state $\rho \in H_1\otimes H_2\otimes\cdots\otimes H_N$, $k$-nonseparability measures $A^k_{q,s}(\rho)$ and $G^k_{q,s}(\rho)$ satisfy,
\begin{equation}\label{r1}
\begin{array}{rl}
A^k_{q_1,s}(\rho)\geq A^k_{q_2,s}(\rho)
\end{array}
\end{equation}
and
\begin{equation}\label{r2}
\begin{array}{rl}
G^k_{q_1,s}(\rho)\geq G^k_{q_2,s}(\rho)
\end{array}
\end{equation}
for $(q_1,s),(q_2,s)\in\mathcal C$
with $0<q_1\le q_2$,
$q_1,q_2\neq1$,
and either $s\ge1$ or $s=0$,
where the case $s=0$ is understood as the limiting R\'enyi entropy case.

\emph{Proof.} We first prove that inequality (\ref{r1}) holds for any $N$-partite pure state $|\psi\rangle$.
Assuming that $\gamma=\{\gamma_1,\gamma_2,\cdots,\gamma_k\}$ is the optimal $k$-partition satisfying $A^k_{q_1,s}(|\psi\rangle)=\dfrac{\sum\limits_{t=1}^{k}S_{q_1,s}(\rho_{\gamma_t})}{k}$, we obtain
\begin{equation*}
\begin{array}{rl}
A^k_{q_1,s}(|\psi\rangle)=&\dfrac{\sum\limits_{t=1}^{k}S_{q_1,s}(\rho_{\gamma_t})}{k}\\
\geq&\dfrac{\sum\limits_{t=1}^{k}S_{q_2,s}(\rho_{\gamma_t})}{k}\\
\geq& A^k_{q_2,s}(|\psi\rangle),
\end{array}
\end{equation*}
The first inequality follows from the monotonicity of the unified
$(q,s)$-entropy with respect to $q$,
\begin{equation}\label{r3}
S_{q_1,s}(\sigma)
\geq
S_{q_2,s}(\sigma),
\end{equation}
valid for $0<q_1\le q_2$, $q_1,q_2\neq1$,
and either $s\ge1$ or $s=0$ \cite{LuoPRA2025},
where $s=0$ corresponds to the R\'enyi entropy obtained by continuous extension.

We now turn to inequality (\ref{r1}) for any $N$-partite mixed state $\rho$. Let $\{p_l,|\psi^{(l)}\rangle\}$ be the optimal pure state decomposition in the definition of $A^k_{q_1,s}(\rho)$, then
\begin{equation*}
\begin{array}{rl}
A^k_{q_1,s}(\rho)=&\sum\limits_{l}p_lA^k_{q_1,s}(|\psi^{(l)}\rangle)\\
\geq&\sum\limits_{l}p_lA^k_{q_2,s}(|\psi^{(l)}\rangle)\\
\geq&A^k_{q_2,s}(\rho),
\end{array}
\end{equation*}
where the first inequality follows from inequality (\ref{r1}) holds for any $N$-partite pure state.

Similarly, we can also prove inequality (\ref{r2}) by inequality (\ref{r3}). \hfill$\blacksquare$

\emph{Theorem 8.} For any $N$-partite quantum state $\rho \in H_1\otimes H_2\otimes\cdots\otimes H_N$, $k$-nonseparability measures $A^k_{q,s}(\rho)$ and $G^k_{q,s}(\rho)$ satisfy,
\begin{equation}\label{r1-1}
\begin{array}{rl}
A^k_{q,s_1}(\rho)\leq A^k_{q,s_2}(\rho)
\end{array}
\end{equation}
and
\begin{equation}\label{r2-1}
\begin{array}{rl}
G^k_{q,s_1}(\rho)\leq G^k_{q,s_2}(\rho)
\end{array}
\end{equation}
for $0<s_1\leq s_2$ with $0<q<1$; and
\begin{equation}\label{r1-2}
\begin{array}{rl}
A^k_{q,s_1}(\rho)\geq A^k_{q,s_2}(\rho)
\end{array}
\end{equation}
and
\begin{equation}\label{r2-2}
\begin{array}{rl}
G^k_{q,s_1}(\rho)\geq G^k_{q,s_2}(\rho)
\end{array}
\end{equation}
for $0<s_1\leq s_2$ with $q>1$.

\emph{Proof.} We first demonstrate that inequality (\ref{r1-1}) holds  for  any $N$-partite quantum state. For an $N$-partite pure state $|\psi\rangle$, let $\gamma=\{\gamma_1,\gamma_2,\cdots,\gamma_k\}$ be the optimal $k$-partition such that $A^k_{q,s_2}(|\psi\rangle)=\dfrac{\sum\limits_{t=1}^{k}S_{q,s_2}(\rho_{\gamma_t})}{k}$. Then
\begin{equation*}
\begin{array}{rl}
A^k_{q,s_2}(|\psi\rangle)=&\dfrac{\sum\limits_{t=1}^{k}S_{q,s_2}(\rho_{\gamma_t})}{k}\\
\geq&\dfrac{\sum\limits_{t=1}^{k}S_{q,s_1}(\rho_{\gamma_t})}{k}\\
\geq& A^k_{q,s_1}(|\psi\rangle),
\end{array}
\end{equation*}
where the first inequality is obtained by the inequality
\begin{equation}\label{r3-1}
\begin{array}{rl}
S_{q,s_1}(\rho_{\gamma_t})\leq S_{q,s_2}(\rho_{\gamma_t})
\end{array}
\end{equation}
for $0<s_1\leq s_2$ with $0<q<1$. The proof of inequality (\ref{r3-1}) is provided in Appendix G.
For an $N$-partite mixed state $\rho$, let $\{p_l,|\psi^{(l)}\rangle\}$ be the optimal pure state decomposition in the definition of $A^k_{q,s_2}(\rho)$, then
\begin{equation*}
\begin{array}{rl}
A^k_{q,s_2}(\rho)=&\sum\limits_{l}p_lA^k_{q,s_2}(|\psi^{(l)}\rangle)\\
\geq&\sum\limits_{l}p_lA^k_{q,s_1}(|\psi^{(l)}\rangle)\\
\geq&A^k_{q,s_1}(\rho),
\end{array}
\end{equation*}
where the first inequality holds because inequality (\ref{r1-1}) has been established for arbitrary pure states.

We now prove inequality~(\ref{r1-2}).
Let $\eta=\{\eta_1,\eta_2,\cdots,\eta_k\}$ be the optimal $k$-partition such that $A^k_{q,s_1}(|\psi\rangle)=\dfrac{\sum\limits_{t=1}^{k}S_{q,s_1}(\rho_{\eta_t})}{k}$. Then we can obtain
\begin{equation*}
\begin{array}{rl}
A^k_{q,s_1}(|\psi\rangle)=&\dfrac{\sum\limits_{t=1}^{k}S_{q,s_1}(\rho_{\eta_t})}{k}\\
\geq&\dfrac{\sum\limits_{t=1}^{k}S_{q,s_2}(\rho_{\eta_t})}{k}\\
\geq& A^k_{q,s_2}(|\psi\rangle),
\end{array}
\end{equation*}
where the first inequality is obtained by the inequality
\begin{equation}\label{r3-2}
\begin{array}{rl}
S_{q,s_1}(\rho_{\eta_t})\geq S_{q,s_2}(\rho_{\eta_t})
\end{array}
\end{equation}
for $0<s_1\leq s_2$ with $q>1$. The proof of inequality (\ref{r3-2}) is provided in Appendix G.
For any $N$-partite mixed state $\rho$, let $\{p'_m,|\varphi^{(m)}\rangle\}$ be the optimal pure state decomposition in the definition of $A^k_{q,s_1}(\rho)$, then
\begin{equation*}
\begin{array}{rl}
A^k_{q,s_1}(\rho)=&\sum\limits_{m}p'_mA^k_{q,s_1}(|\varphi^{(m)}\rangle)\\
\geq&\sum\limits_{m}p'_mA^k_{q,s_2}(|\varphi^{(m)}\rangle)\\
\geq&A^k_{q,s_2}(\rho),
\end{array}
\end{equation*}
where the first inequality holds because when $0<s_1\leq s_2$ with $q>1$, inequality (\ref{r1-2}) is satisfied for an arbitrary $N$-partite pure state.

Inequalities (\ref{r2-1}) and (\ref{r2-2}) can be proved analogously using inequalities (\ref{r3-1}) and (\ref{r3-2}), respectively. \hfill$\blacksquare$

We next discuss the ordering of these two $k$-nonseparability measures $A^k_{q,s}(\rho)$ and $G^k_{q,s}(\rho)$ for arbitrary pure states, with the parameters $q$ and $s$ fixed.

\emph{Theorem 9.}
Let $|\psi\rangle$ and $|\widetilde{\psi}\rangle$ be two pure states on $H_1\otimes H_2\otimes\cdots\otimes H_N$. For every nonempty proper subset
$\Lambda\subset\{1,2,\ldots,N\}$, let $\rho_\Lambda$ and $\widetilde{\rho}_\Lambda$ denote the corresponding reduced density matrices.

If
\begin{equation}
\boldsymbol{\lambda}(\rho_\Lambda)
\prec
\boldsymbol{\lambda}(\widetilde{\rho}_\Lambda)
\end{equation}
for every nonempty proper subset $\Lambda$, where $\boldsymbol{\lambda}(X)$ denotes the eigenvalue vector of $X$ arranged in nonincreasing order, then
\begin{equation}\label{r1-3}
A^k_{q,s}(|\psi\rangle)
\geq
A^k_{q,s}(|\widetilde{\psi}\rangle)
\end{equation}
and
\begin{equation}\label{r2-3}
G^k_{q,s}(|\psi\rangle)
\geq
G^k_{q,s}(|\widetilde{\psi}\rangle)
\end{equation}
for $q>0$, $q\neq1$, and $s>0$.

\emph{Proof.} The case $q=1$ follows from the von Neumann entropy. We first show that inequality~(\ref{r1-3}) holds. Let $\gamma=\{\gamma_1,\gamma_2,\cdots,\gamma_k\}$ be the optimal $k$-partition such that $A^k_{q,s}(|\psi\rangle)=\dfrac{\sum\limits_{t=1}^{k}S_{q,s}(\rho_{\gamma_t})}{k}$, then one obtains
\begin{equation*}
\begin{array}{rl}
A^k_{q,s}(|\psi\rangle)=&\dfrac{\sum\limits_{t=1}^{k}S_{q,s}(\rho_{\gamma_t})}{k}\\
\geq&\dfrac{\sum\limits_{t=1}^{k}S_{q,s}(\widetilde{\rho}_{\gamma_t})}{k}\\
\geq& A^k_{q,s}(|\widetilde{\psi}\rangle),
\end{array}
\end{equation*}
where $\rho_{\gamma_t}$ and $\widetilde{\rho}_{\gamma_t}$ are the reduced density matrices of the pure states $|\psi\rangle$ and $|\widetilde{\psi}\rangle$ with respect to the subsystem $\bigotimes\limits_{i\in \gamma_t}H_i$, respectively.
Here the first inequality is true because of  the inequality
\begin{equation}\label{r3-3}
\begin{array}{rl}
S_{q,s}(\rho_{\gamma_t})\geq S_{q,s}(\widetilde{\rho}_{\gamma_t})
\end{array}
\end{equation}
for $q>0$, $q\neq1$, $s>0$ and $\lambda_{\psi_{\gamma_t}}\prec\lambda_{\widetilde{\psi}_{\gamma_t}}$. A proof of inequality (\ref{r3-3}) is provided in Appendix H.

Similarly, using inequality (\ref{r3-3}), we obtain that inequality (\ref{r2-3}) holds. \hfill$\blacksquare$

\section{Conclusion}
In this work, we introduce a bipartite entanglement measure $E_{q,s}^{A|B}(\rho)$ based on unified $(q,s)$-entropy and derive a series of lower bounds for it via local orthonormal observables. Concrete examples demonstrate that these lower bounds are tighter than several existing results. For multipartite systems, we introduce two distinct measures $A_{q,s}^{k}(\rho)$ and $G_{q,s}^{k}(\rho)$, both based on unified $(q,s)$-entropy, to characterize multipartite entanglement in terms of $k$-nonseparability. We prove that these two entanglement measures satisfy faithfulness, invariance under local unitary transformations, and convexity. When $(q,s)\in\mathcal{C}$, we further prove that these two $k$-nonseparability measures are monotonic and strongly monotonic under LOCC.

In addition, we establish several ordering relations for $A_{q,s}^{k}$ and $G_{q,s}^{k}$. We show that both measures exhibit monotonic behavior with respect to the parameters $q$ and $s$ in the corresponding parameter regimes. Furthermore, for pure states, we prove that both measures are compatible with the majorization ordering of the spectra of all reduced density operators. These results provide a unified framework for analyzing multipartite entanglement.

\appendix
\section{The proof of Proposition 1}
For any bipartite quantum state $|\psi\rangle\in H_A\otimes H_B$ with $H_A= H_B=H$ and $\dim H=d$, there exist orthonormal
bases $\{|f_m\rangle:m=1,2,\cdots,d\}$ in $H_A$ and $\{|h_l\rangle:l=1,2,\cdots,d\}$ in $H_B$ such that
$|\psi\rangle=\sum\limits_{m=1}^{d}\lambda_m|f_m\rangle|h_m\rangle$,
Thus, we have
\begin{equation}\label{L1}
\begin{array}{rl}
L(|\psi\rangle\langle\psi|)=&\sum\limits_{u,v=1}^{d^2}\langle\psi|B_u\otimes B_v|\psi\rangle|e_u\rangle\langle e_v|\\
=&\sum\limits_{u,v=1}^{d^2}(\sum\limits_{m=1}^{d}\lambda_m\langle f_m|\langle h_m|)B_u\otimes B_v(\sum\limits_{l=1}^{d}\lambda_l|f_l\rangle|h_l\rangle)|e_u\rangle\langle e_v|\\
=&\sum\limits_{m=1}^{d}\sum\limits_{l=1}^{d}\lambda_m\lambda_l\Big(\sum\limits_{u=1}^{d^2}
\langle f_m|B_u|f_l\rangle|e_u\rangle\Big)\Big(\sum\limits_{v=1}^{d^2}\langle h_m|B_v|h_l\rangle
\langle e_v|\Big)\\
=&\sum\limits_{m=1}^{d}\sum\limits_{l=1}^{d}\lambda_m\lambda_l|\eta_{ml}\rangle\langle\tau_{ml}|,
\end{array}
\end{equation}
where $|\eta_{ml}\rangle=\sum\limits_{u=1}^{d^2}
\langle f_m|B_u|f_l\rangle|e_u\rangle$ and $|\tau_{ml}\rangle=\sum\limits_{v=1}^{d^2}\langle h_m|B_v|h_l\rangle|e_v\rangle$.

Since both $\{B_u\}_{u=1}^{d^2}$ and
$\{B'_v\}_{v=1}^{d^2}$ are orthonormal bases of the real Hilbert
space of Hermitian operators equipped with the Hilbert--Schmidt
inner product, there exists a real orthogonal matrix
$O=(b_{uv})$ such that
\begin{equation*}
B_u=\sum_{v=1}^{d^2}b_{uv}B'_v,
\end{equation*}
and
\begin{equation*}
\sum_{u=1}^{d^2}b_{uv}b_{uv'}
=\delta_{vv'}.
\end{equation*}
Then,
\begin{equation*}
\begin{array}{rl}
\sum\limits_{u=1}^{d^2}B_u\otimes B_u=&\sum\limits_{u=1}^{d^2}\Big(\sum\limits_{v=1}^{d^2}b_{uv}B'_v\Big)\otimes\Big(\sum\limits_{v'=1}^{d^2}b_{uv'}B'_{v'}\Big)\\
=&\sum\limits_{v,v'=1}^{d^2}\Big(\sum\limits_{u=1}^{d^2}b_{uv}b_{uv'}\Big)B'_v\otimes B'_{v'}\\
=&\sum\limits_{v=1}^{d^2}B'_v\otimes B'_{v}.
\end{array}
\end{equation*}
Consequently, $\sum\limits_{u=1}^{d^2}B_u\otimes B_u$ is independent of the choice of the orthonormal observable basis.
In particular, we take $\{B_u:u=1,2,\cdots,d^2\}=\{E_m,E^+_{ml},E^-_{ml}\}$ with
\begin{equation*}
\begin{array}{rl}
E_m=&|f_m\rangle\langle f_m|,\quad m=1,2,\cdots,d,\\\\
E^+_{ml}=&\dfrac{1}{\sqrt{2}}(|f_m\rangle\langle f_l|+|f_l\rangle\langle f_m|), \quad m<l\textrm{ and  }m,l =1,2,\cdots,d,\\\\
E^-_{ml}=&\dfrac{i}{\sqrt{2}}(|f_m\rangle\langle f_l|-|f_l\rangle\langle f_m|), \quad m<l\textrm{ and  }m,l =1,2,\cdots,d.
\end{array}
\end{equation*}
The operators $\{E_m,E^+_{ml},E^-_{ml}\}$ form an orthonormal Hermitian operator basis.

Then, we have
\begin{equation}\label{16-1}
\begin{array}{rl}
\sum\limits_{u=1}^{d^2}B_u\otimes B_u
=&\sum\limits_{m=1}^{d}|f_m\rangle\langle f_m|\otimes |f_m\rangle\langle f_m|\\
&+\dfrac{1}{2}\sum\limits_{m<l}(|f_m\rangle\langle f_l|+|f_l\rangle\langle f_m|)\otimes(|f_m\rangle\langle f_l|+|f_l\rangle\langle f_m|)\\
&-\dfrac{1}{2}\sum\limits_{m<l}(|f_m\rangle\langle f_l|-|f_l\rangle\langle f_m|)\otimes(|f_m\rangle\langle f_l|-|f_l\rangle\langle f_m|)\\
=&\sum\limits_{m=1}^{d}|f_mf_m\rangle\langle f_mf_m|+\sum\limits_{m<l}(|f_mf_l\rangle\langle f_lf_m|+|f_lf_m\rangle\langle f_mf_l|)\\
=&\sum\limits_{m,l=1}^d|f_mf_l\rangle\langle f_lf_m|.
\end{array}
\end{equation}
Using Eq. (\ref{16-1}), we obtain
\begin{equation}\label{16-3}
\begin{array}{rl}
\langle\eta_{ml}|\eta_{m'l'}\rangle=&\sum\limits_{u,v=1}^{d^2}
\langle f_l|B_u|f_m\rangle\langle f_{m'}|B_v|f_{l'}\rangle\langle e_u|e_v\rangle\\
=&\langle f_l f_{m'}|\Big(\sum\limits_{u=1}^{d^2}B_u\otimes B_u\Big)|f_m f_{l'}\rangle\\
=&\langle f_l f_{m'}|\Big(\sum\limits_{m,l=1}^d|f_mf_l\rangle\langle f_lf_m|\Big)|f_m f_{l'}\rangle\\
=&\delta_{ll'}\delta_{mm'}.
\end{array}
\end{equation}

(1) Let us first consider the special case where $|\eta_{ml}\rangle=|\tau_{ml}\rangle$ for any $m,l=1,2,\cdots,d$.
It follows immediately from Eq. (\ref{L1}) that $L(|\psi\rangle\langle\psi|)$ is positive semidefinite, and
\begin{equation*}
\begin{array}{rl}
\|L(|\psi\rangle\langle\psi|)\|_{\operatorname{Tr}}
=&\operatorname{Tr}\Big(\sum\limits_{m=1}^{d}\sum\limits_{l=1}^{d}\lambda_m\lambda_l|\eta_{ml}\rangle\langle\eta_{ml}|\Big)\\
=&\Big(\sum\limits_{m=1}^{d}\lambda_m\Big)^2,
\end{array}
\end{equation*}
where second equation holds by Eq. (\ref{16-3}).

(2) Let us consider the general case.
Similar to the proof of Eq. (\ref{16-3}), we can derive
\begin{equation}\label{general1}
\begin{array}{rl}
\langle\tau_{ml}|\tau_{m'l'}\rangle
=&\Big(\sum\limits_{v=1}^{d^2}\langle h_{m}|B_v|h_{l}\rangle\langle e_{v}|\Big)\Big(\sum\limits_{v'=1}^{d^2}\langle h_{l'}|B_{v'}|h_{m'}\rangle|e_{v'}\rangle\Big)\\
=&\sum\limits_{v=1}^{d^2}\langle h_{m}|B_v|h_{l}\rangle\langle h_{l'}|B_{v}|h_{m'}\rangle\\
=&\langle h_m h_{l'}|\Big(\sum\limits_{v=1}^{d^2}B_v\otimes B_v\Big)|h_l h_{m'}\rangle\\
=&\langle h_m h_{l'}|\Big(\sum\limits_{m,l=1}^d|h_mh_l\rangle\langle h_lh_m|\Big)|h_l h_{m'}\rangle\\
=&\delta_{ll'}\delta_{mm'}.
\end{array}
\end{equation}
From  Eqs. (\ref{16-3}) and (\ref{general1}), we have
\begin{equation*}\label{general2}
\begin{array}{rl}
\langle\eta_{ml}|\eta_{m'l'}\rangle=\langle\tau_{ml}|\tau_{m'l'}\rangle=\delta_{ll'}\delta_{mm'}.
\end{array}
\end{equation*}

Since both sets $\{|\eta_{ml}\rangle\}_{m,l=1}^{d}$ and $\{|\tau_{ml}\rangle\}_{m,l=1}^{d}$ are orthonormal bases of the same Hilbert space $\mathbb C^{d^2}$, there exists a unitary operator $U:\mathbb C^{d^2}\rightarrow\mathbb C^{d^2}$ such that
\begin{equation}\label{general3}
\begin{array}{rl}
|\tau_{ml}\rangle=U|\eta_{ml}\rangle,
\end{array}
\end{equation}
for any $m,l=1,2,\cdots,d$ \cite{WangFeiPRA2025}. By Eqs. (\ref{L1}) and (\ref{general3}), one has
\begin{equation*}\label{L2}
\begin{array}{rl}
\|L(|\psi\rangle\langle\psi|)\|_{\operatorname{Tr}}
=&\|\sum\limits_{m=1}^{d}\sum\limits_{l=1}^{d}\lambda_m\lambda_l|\eta_{ml}\rangle\langle\tau_{ml}|\|_{\operatorname{Tr}}\\
=&\|\Big(\sum\limits_{m=1}^{d}\sum\limits_{l=1}^{d}\lambda_m\lambda_l|\eta_{ml}\rangle\langle\eta_{ml}|\Big)U^\dagger\|_{\operatorname{Tr}}\\
=&\|\sum\limits_{m=1}^{d}\sum\limits_{l=1}^{d}\lambda_m\lambda_l|\eta_{ml}\rangle\langle\eta_{ml}|\|_{\operatorname{Tr}}\\
=&\Big(\sum\limits_{m=1}^{d}\lambda_m\Big)^2.
\end{array}
\end{equation*}
where the last equality follows from the trace norm formula of the partial transpose of a pure bipartite state.

Combining the above two cases, we obtain
\begin{equation*}\label{L3}
\begin{array}{rl}
\|L(|\psi\rangle\langle\psi|)\|_{\operatorname{Tr}}=&\Big(\sum\limits_{m=1}^{d}\lambda_m\Big)^2=\|\rho^{T_A}\|_{\operatorname{Tr}},
\end{array}
\end{equation*}
for any pure state $\rho=|\psi\rangle\langle\psi|\in H_A\otimes H_B$ with $H_A= H_B=H$ and dim $H=d$.

\section{The proof of Theorem 2-4  }
\subsection{The proof of Theorem 2  }
For any 2-qudit quantum pure state $|\psi\rangle$,
\begin{equation*}
\begin{array}{rl}
E_{q,s}^{A|B}(|\psi\rangle)=&S_{q,s}(\rho_A)\\
=&\dfrac{1-(\operatorname{Tr}\rho_A^q)^s}{(q-1)s}\\
\geq&\dfrac{1-\left[1-\dfrac{1}{d^{2q-2}-d^{q-1}}\Big(\|L(|\psi\rangle\langle\psi|)\|_{\operatorname{Tr}}^{q-1}-1\Big)^2\right]^s}{(q-1)s}
\end{array}
\end{equation*}
for $q\geq2$ and $0<s\leq1$.
Here the inequality follows from Eq. (\ref{L3-0}), inequality (\ref{Cbound1}), and the monotonicity of $y=x^s$ with respect to $x$.

For any 2-qudit mixed state $\rho$, suppose that $\{p_l,|\psi^{(l)}\rangle\}$ is the optimal decomposition, then we obtain
\begin{equation*}
\begin{array}{rl}
E_{q,s}^{A|B}(\rho)=&\sum\limits_lp_lE_{q,s}^{A|B}(|\psi^{(l)}\rangle)\\
\geq&\sum\limits_lp_l\dfrac{1-\left[1-\dfrac{1}{d^{2q-2}-d^{q-1}}\Big(\|L(|\psi^{(l)}\rangle\langle\psi^{(l)}|)\|_{\operatorname{Tr}}^{q-1}-1\Big)^2\right]^s}{(q-1)s}\\
\geq&\dfrac{1-\left[1-\sum\limits_lp_l\dfrac{1}{d^{2q-2}-d^{q-1}}\Big(\|L(|\psi^{(l)}\rangle\langle\psi^{(l)}|)\|_{\operatorname{Tr}}^{q-1}-1\Big)^2\right]^s}{(q-1)s}\\
\geq&\dfrac{1-\left[1-\dfrac{1}{d^{2q-2}-d^{q-1}}\Big(\sum\limits_lp_l\|L(|\psi^{(l)}\rangle\langle\psi^{(l)}|)\|_{\operatorname{Tr}}^{q-1}-1\Big)^2\right]^s}{(q-1)s}\\
\geq&\dfrac{1-\left[1-\dfrac{1}{d^{2q-2}-d^{q-1}}\Big(\big(\sum\limits_lp_l\|L(|\psi^{(l)}\rangle\langle\psi^{(l)}|)\|_{\operatorname{Tr}}\big)^{q-1}-1\Big)^2\right]^s}{(q-1)s}\\
\geq&\dfrac{1-\left[1-\dfrac{1}{d^{2q-2}-d^{q-1}}\Big(\|L(\rho)\|_{\operatorname{Tr}}^{q-1}-1\Big)^2\right]^s}{(q-1)s}
\end{array}
\end{equation*}
for $q\geq2$ and $0<s\leq1$. The second inequality follows from the concavity of
$f(x)=(1-x)^s$ for $0<s\le1$.
The third inequality follows from Jensen's inequality for the convex function
$x^2$, namely, $\sum_l p_lx_l^2\geq(\sum_lp_lx_l)^2 $.
The fourth inequality follows from Jensen's inequality applied to the convex function
$x^{q-1}$ for $q\ge2$.
Finally,
\[
\|L(\rho)\|_{\operatorname{Tr}}
\leq
\sum_l p_l
\|L(|\psi^{(l)}\rangle\langle\psi^{(l)}|)\|_{\operatorname{Tr}}
\]
follows from the linearity of $L$ and the convexity of the trace norm.

\subsection{The proof of Theorem 3  }
For any 2-qudit quantum pure state $|\psi\rangle$,
\begin{equation*}
\begin{array}{rl}
E_{q,s}^{A|B}(|\psi\rangle)=&S_{q,s}(\rho_A)\\
=&\dfrac{1-(\operatorname{Tr}\rho_A^q)^s}{(q-1)s}\\
\geq&\dfrac{1-\left[1-\dfrac{1-2^{1-q}}{2-2^{2-r}}\Big(\|L(|\psi\rangle\langle\psi|)\|_{\operatorname{Tr}}-1\Big)^2\right]^s}{(q-1)s}
\end{array}
\end{equation*}
for $0<s\leq1$, $d=2$ with $2.4721=r\leq q<3$. Here the inequality holds because of Eq. (\ref{L3-0}),inequality (\ref{Cbound2}) and the fact that function $y=x^s$ is monotonically increasing in $x$.

For any 2-qudit mixed state $\rho$, suppose that $\{p_l,|\psi^{(l)}\rangle\}$ is the optimal decomposition, then
\begin{equation*}
\begin{array}{rl}
E_{q,s}^{A|B}(\rho)=&\sum\limits_lp_lE_{q,s}^{A|B}(|\psi^{(l)}\rangle)\\
\geq&\sum\limits_lp_l\dfrac{1-\left[1-\dfrac{1-2^{1-q}}{2-2^{2-r}}\Big(\|L(|\psi^{(l)}\rangle\langle\psi^{(l)}|)\|_{\operatorname{Tr}}-1\Big)^2\right]^s}{(q-1)s}\\
\geq&\dfrac{1-\left[1-\sum\limits_lp_l\dfrac{1-2^{1-q}}{2-2^{2-r}}\Big(\|L(|\psi^{(l)}\rangle\langle\psi^{(l)}|)\|_{\operatorname{Tr}}-1\Big)^2\right]^s}{(q-1)s}\\
\geq&\dfrac{1-\left[1-\dfrac{1-2^{1-q}}{2-2^{2-r}}\Big(\sum\limits_lp_l\|L(|\psi^{(l)}\rangle\langle\psi^{(l)}|)\|_{\operatorname{Tr}}-1\Big)^2\right]^s}{(q-1)s}\\
\geq&\dfrac{1-\left[1-\dfrac{1-2^{1-q}}{2-2^{2-r}}\Big(\|L(\rho)\|_{\operatorname{Tr}}-1\Big)^2\right]^s}{(q-1)s}
\end{array}
\end{equation*}
for $0<s\leq1$, $d=2$ with $2.4721=r\leq q<3$.
Here the second and third inequalities are obtained, respectively, from the concavity of $y=x^s$ for $0<s\leq1$ and the convexity of $y=x^2$.
The last inequality follows from the convexity of the trace norm and the linearity of the function $L(\rho)$.
Hence, inequality (\ref{Ebound2-0}) holds for $0<s\leq1$, $d=2$ with $2.4721=r\leq q<3$. Similarly, inequality (\ref{Ebound3-0}) can be shown to hold for $0<s\leq1$, either $d=2$ with $q\geq3$  or $d\geq3$ with $q\geq2$.

\subsection{The proof of Theorem 4  }
For any 2-qudit quantum pure state $|\psi\rangle$,
\begin{equation*}
\begin{array}{rl}
E_{q,s}^{A|B}(|\psi\rangle)=&S_{q,s}(\rho_A)\\
=&\dfrac{(\operatorname{Tr}\rho_A^q)^s-1}{(1-q)s}\\
\geq&\dfrac{\left[\dfrac{d^{1-q}-1}{d-1}\big(\|L(|\psi\rangle\langle\psi|)\|_{\operatorname{Tr}}-1\big)+1\right]^{s}-1}{(1-q)s}
\end{array}
\end{equation*}
for $0<q\leq\frac{1}{2}$ with $s\geq1$,
Here the inequality  follows from Eq. (\ref{L3-0}), inequality (\ref{Cbound5}) and the monotonically increasing property  of $y=x^{s}$ in $x$.

For any 2-qudit mixed state $\rho$, suppose that $\{p_l,|\psi^{(l)}\rangle\}$ is the optimal decomposition, then
\begin{equation*}
\begin{array}{rl}
E_{q,s}^{A|B}(\rho)=&\sum\limits_lp_lE_{q,s}^{A|B}(|\psi^{(l)}\rangle)\\
\geq&\sum\limits_lp_l\dfrac{\left[\dfrac{d^{1-q}-1}{d-1}\big(\|L(|\psi^{(l)}\rangle\langle\psi^{(l)}|)\|_{\operatorname{Tr}}-1\big)+1\right]^{s}-1}{(1-q)s}\\
\geq&\dfrac{\left[\dfrac{d^{1-q}-1}{d-1}\big(\sum\limits_lp_l\|L(|\psi^{(l)}\rangle\langle\psi^{(l)}|)\|_{\operatorname{Tr}}-1\big)+1\right]^{s}-1}{(1-q)s}\\
\geq&\dfrac{\left[\dfrac{d^{1-q}-1}{d-1}\big(\|L(\rho)\|_{\operatorname{Tr}}-1\big)+1\right]^{s}-1}{(1-q)s}
\end{array}
\end{equation*}
for $0<q\leq\frac{1}{2}$ with $s\geq1$.
Since $a_l\ge1$, the convexity of $x^s$ for $s\ge1$
implies Jensen's inequality.
The last inequality follows from the convexity of the trace norm and the linearity of function $L(\rho)$.

\paragraph{Notation for LOCC operations.}
Throughout the following proofs, an LOCC instrument admits a Kraus representation
$\{M_l\}$ generated by a finite sequence of local measurements and
classical communication.
The corresponding quantum operation is
\begin{equation}
\Lambda_{\rm LOCC}(\rho)
=
\sum_l M_l\rho M_l^\dagger ,
\end{equation}
where
\begin{equation}
\sum_l M_l^\dagger M_l=I .
\end{equation}
For an input pure state $|\psi\rangle$, the state corresponding to the
$ l$-th measurement outcome is
\begin{equation}
|\omega^{(l)}\rangle
=
\frac{M_l|\psi\rangle}{\sqrt{p_l}},
\end{equation}
where
\begin{equation}
p_l=
\langle\psi|M_l^\dagger M_l|\psi\rangle .
\end{equation}
Hence,
\begin{equation}
\Lambda_{\rm LOCC}
(|\psi\rangle\langle\psi|)
=
\sum_l p_l
|\omega^{(l)}\rangle
\langle\omega^{(l)}|.
\end{equation}

\section{Proof of inequality (\ref{iva4})}

Consider the bipartition $\gamma_t|\bar{\gamma}_t$.
The parties contained in $\gamma_t$ and $\bar{\gamma}_t$ can be regarded
as two composite subsystems. Therefore, any multipartite LOCC protocol
induces a bipartite LOCC operation with respect to this bipartition,
and Theorem~1 applies to
$E_{q,s}^{\gamma_t|\bar{\gamma}_t}$.

If the deterministic LOCC channel
$\Lambda_{\rm LOCC}$ maps the pure state
$|\psi\rangle$
to the pure state
$|\widetilde{\psi}\rangle$. For a deterministic LOCC transformation,
the monotonicity condition becomes
\begin{equation}
E_{q,s}^{\gamma_t|\bar{\gamma}_t}(|\psi\rangle)
\geq
E_{q,s}^{\gamma_t|\bar{\gamma}_t}(|\widetilde{\psi}\rangle).
\end{equation}
Using Definition~1, this is equivalent to
\begin{equation}
S_{q,s}(\rho_{\gamma_t})
\geq
S_{q,s}(\widetilde{\rho}_{\gamma_t}),
\end{equation}
which proves inequality~(\ref{iva4}) for $(q,s)\in\mathcal{C}$.

\section{The proof of inequality (\ref{iva0})  }

Let $\gamma_t$ be any nonempty proper subset of $\{1,2,\ldots,N\}$, and let $\bar{\gamma}_t=\{1,2,\ldots,N\}\setminus\gamma_t$. We regard the multipartite state $\rho$ as a bipartite state with respect to the bipartition $\gamma_t|\bar{\gamma}_t$, namely,
\[
H_A=\bigotimes_{i\in\gamma_t}H_i,
\qquad
H_B=\bigotimes_{i\in\bar{\gamma}_t}H_i,
\]
where the two parties are regarded as
the composite systems
$H_{\gamma_t}$ and $H_{\bar{\gamma}_t}$. Therefore, Theorem~1 can be applied to $E_{q,s}^{\gamma_t|\bar{\gamma}_t}$. That is,
\begin{equation}\label{iva00}
E_{q,s}^{\gamma_t|\bar{\gamma}_t}(\rho)\geq \sum\limits_lp_lE_{q,s}^{\gamma_t|\bar{\gamma}_t}(\sigma^{(l)}),
\end{equation}
 where $\sigma^{(l)}$ is obtained with probability $p_l$ by applying LOCC $\Lambda_{\rm LOCC}$ to $\rho$. When $\rho=|\psi\rangle\langle\psi|$ is a pure state, we obtain
\begin{equation*}
\sum\limits_lp_lE_{q,s}^{\gamma_t|\bar{\gamma}_t}(|\omega^{(l)}\rangle)
\leq E_{q,s}^{\gamma_t|\bar{\gamma}_t}(|\psi\rangle).
\end{equation*}

\section{ The proof of item (iv) for $G^k_{q,s}(\rho)$ }

For any pure state $|\psi\rangle$,  we will consider two cases: one in which $\Lambda_{\rm LOCC}(|\psi\rangle\langle\psi|)$ is a pure state, and one in which $\Lambda_{\rm LOCC}(|\psi\rangle\langle\psi|)$ is a mixed state.

Case 1. If $\Lambda_{\rm LOCC}(|\psi\rangle\langle\psi|)$ is a pure state, let $\rho=|\psi\rangle\langle\psi|, \rho_{\gamma_t}=\operatorname{Tr}_{\bar{\gamma}_t}(\rho)$,  $\widetilde{\rho}=|\widetilde{\psi}\rangle\langle\widetilde{\psi}|$ and $\widetilde{\rho}_{\gamma_t}=\operatorname{Tr}_{\bar{\gamma}_t}(\widetilde{\rho})$ with
$|\widetilde{\psi}\rangle=\Lambda_{\rm LOCC}(|\psi\rangle\langle\psi|)$ for the deterministic pure-state output. Then one has
\begin{equation*}
\begin{array}{rl}
G^k_{q,s}[\Lambda_{\rm LOCC}(|\psi\rangle\langle\psi|)]
=&\Big(\prod\limits_{\gamma\in P_k}\big[\sum\limits_{t=1}^{k}S_{q,s}(\widetilde{\rho}_{\gamma_t})/k\big]\Big)^{\frac{1}{c(k)}}\\
\leq&\Big(\prod\limits_{\gamma\in P_k}\big[\sum\limits_{t=1}^{k}S_{q,s}(\rho_{\gamma_t})/k\big]\Big)^{\frac{1}{c(k)}}\\
=&G^k_{q,s}(|\psi\rangle),
\end{array}
\end{equation*}
where the inequality  holds by inequality (\ref{iva4}).

Case 2. If the output state $\Lambda_{\rm LOCC}(|\psi\rangle\langle\psi|)$ is mixed, consider an LOCC instrument described by Kraus operators
$\{M_l\}$ with $\sum\limits_lM_l^\dagger M_l=I. $ Then
$\Lambda_{\rm LOCC}(|\psi\rangle\langle\psi|)=\sum\limits_lM_l|\psi\rangle\langle\psi|M_l^\dagger=\sum\limits_lp_l|\omega^{(l)}\rangle\langle\omega^{(l)}|$ with $|\omega^{(l)}\rangle=\dfrac{M_l|\psi\rangle}{\sqrt{p_l}}$ and $p_l=\operatorname{Tr}(M_l|\psi\rangle\langle\psi|M_l^\dagger)$.
Using $E_{q,s}^{\gamma_t|\bar{\gamma}_t}(|\psi\rangle)=S_{q,s}(\rho_{\gamma_t}),$
one has
\begin{equation*}
\begin{array}{rl}
G^k_{q,s}[\Lambda_{\rm LOCC}(|\psi\rangle\langle\psi|)]\leq&\sum\limits_lp_lG^k_{q,s}(|\omega^{(l)}\rangle)\\
=&\sum\limits_lp_l\Big(\prod\limits_{\gamma\in P_k}\big[\sum\limits_{t=1}^{k}E_{q,s}^{\gamma_t|\bar{\gamma}_t}(|\omega^{(l)}\rangle)/k\big]\Big)^{\frac{1}{c(k)}}\\
\leq&\Big(\prod\limits_{\gamma\in P_k}\big[\sum\limits_lp_l\sum\limits_{t=1}^{k}E_{q,s}^{\gamma_t|\bar{\gamma}_t}(|\omega^{(l)}\rangle)/k\big]\Big)^{\frac{1}{c(k)}}\\
\leq&\Big(\prod\limits_{\gamma\in P_k}\big[\sum\limits_{t=1}^{k}E_{q,s}^{\gamma_t|\bar{\gamma}_t}(|\psi\rangle)/k\big]\Big)^{\frac{1}{c(k)}}\\
=&G^k_{q,s}(|\psi\rangle).
\end{array}
\end{equation*}
Here the second inequality follows from the fact that the geometric mean function $f=(\prod\limits_{j=1}^mx_j)^{\frac{1}{m}}$ is concave \cite{BoydVandenberghe2004}.
The third  inequality holds by inequality (\ref{iva0}).

Therefore, for any pure state $|\psi\rangle$, one has
\begin{equation}\label{iva2pure}
G^k_{q,s}[\Lambda_{\rm LOCC}(|\psi\rangle\langle\psi|)]\leq G^k_{q,s}(|\psi\rangle).
\end{equation}

Next, let us prove that $G^k_{q,s}(\rho)$ satisfies item (iv) for any mixed state $\rho$.
Suppose that $\{p_l,|\psi^{(l)}\rangle\}$ and $\{p_{il},|\psi^{(il)}\rangle\}$  are the optimal decompositions of $G^k_{q,s}(\rho)$ and $G^k_{q,s}[\Lambda_{\rm LOCC}(|\psi^{(l)}\rangle)]$, respectively.
Then one has
\begin{equation*}
\begin{array}{rl}
G^k_{q,s}(\rho)=&\sum\limits_lp_lG^k_{q,s}(|\psi^{(l)}\rangle)\\
\geq&\sum\limits_lp_lG^k_{q,s}[\Lambda(|\psi^{(l)}\rangle)]\\
=&\sum\limits_lp_l\sum\limits_ip_{il}G^k_{q,s}(|\psi^{(il)}\rangle)\\
\geq&G^k_{q,s}[\Lambda_{\rm LOCC}(\rho)].
\end{array}
\end{equation*}
The first inequality  holds because of  inequality (\ref{iva2pure}).
Using the fact that $\{p_lp_{il},|\psi^{(il)}\rangle\}$ is a pure state ensemble decomposition of $\Lambda_{\rm LOCC}(\rho)$, together with Eq. (\ref{D2-1}), we derive the second inequality.
Hence, for any mixed state $\rho$, we have
\begin{equation*}
G^k_{q,s}[\Lambda_{\rm LOCC}(\rho)]\leq G^k_{q,s}(\rho).
\end{equation*}

\section{Proof of strong monotonicity of $G^k_{q,s}(\rho)$}

We first prove the strong monotonicity of $G^k_{q,s}(\rho)$ under
LOCC operations. It should be noted that for any bipartition
$\gamma_t|\bar{\gamma}_t$,
a multipartite LOCC operation can be regarded as a bipartite LOCC
operation acting on the composite systems
\[
H_{\gamma_t}=\bigotimes_{i\in\gamma_t}H_i,
\qquad
H_{\bar{\gamma}_t}=\bigotimes_{i\in\bar{\gamma}_t}H_i .
\]
Therefore, the strong monotonicity of the bipartite measure
$E_{q,s}^{\gamma_t|\bar{\gamma}_t}$ established in Theorem 1 can be
directly applied.

First, we consider a pure state
$|\psi\rangle$. Suppose that an LOCC instrument
$\Lambda_{\rm LOCC}$
transforms
$|\psi\rangle\langle\psi|$ into a set of pure states $\{|\omega^{(l)}\rangle\}$ with probabilities $\{p_l\}$, where
\[
|\omega^{(l)}\rangle
=
\frac{M_l|\psi\rangle}{\sqrt{p_l}},
\qquad
p_l=
\operatorname{Tr}
(M_l|\psi\rangle\langle\psi|M_l^\dagger).
\]

For any $k$-partition
$\gamma=\{\gamma_1,\gamma_2,\ldots,\gamma_k\}$, we have
\[
E_{q,s}^{\gamma_t|\bar{\gamma}_t}
(|\psi\rangle)
\geq
\sum_l p_l
E_{q,s}^{\gamma_t|\bar{\gamma}_t}
(|\omega^{(l)}\rangle)
\]
according to Theorem 1. Hence,
\begin{equation}\label{iva5-1}
\begin{aligned}
&G^k_{q,s}(|\psi\rangle)\\
=&
\left(
\prod_{\gamma\in P_k}
\left[
\frac{1}{k}
\sum_{t=1}^{k}
E_{q,s}^{\gamma_t|\bar{\gamma}_t}
(|\psi\rangle)
\right]
\right)^{\frac1{c(k)}}\\
\geq&
\left(
\prod_{\gamma\in P_k}
\left[
\frac{1}{k}
\sum_{t=1}^{k}
\sum_l p_l
E_{q,s}^{\gamma_t|\bar{\gamma}_t}
(|\omega^{(l)}\rangle)
\right]
\right)^{\frac1{c(k)}} .
\end{aligned}
\end{equation}

Since all
$E_{q,s}^{\gamma_t|\bar{\gamma}_t}$ are nonnegative, the arguments
of the geometric mean function are nonnegative. The concavity of the
geometric mean function
\[
f(x_1,x_2,\ldots,x_m)
=
(x_1x_2\cdots x_m)^{1/m}
\]
gives
\[
f\left(\sum_l p_lx_l^{(1)},\ldots,
\sum_l p_lx_l^{(m)}\right)
\geq
\sum_l p_l
f(x_l^{(1)},\ldots,x_l^{(m)}).
\]
Therefore,
\begin{equation}\label{iva5-2}
\begin{aligned}
&G^k_{q,s}(|\psi\rangle)\\
\geq&
\sum_l p_l
\left(
\prod_{\gamma\in P_k}
\left[
\frac1k
\sum_{t=1}^{k}
E_{q,s}^{\gamma_t|\bar{\gamma}_t}
(|\omega^{(l)}\rangle)
\right]
\right)^{\frac1{c(k)}}\\
=&
\sum_l p_lG^k_{q,s}(|\omega^{(l)}\rangle).
\end{aligned}
\end{equation}

Thus, $G^k_{q,s}$ is strongly monotonic under LOCC operations for
pure states.

Next, we extend the result to arbitrary mixed states. Let
$\{q_i,|\psi^{(i)}\rangle\}$ be an optimal pure-state decomposition
of $\rho$ satisfying the convex-roof definition of
$G^k_{q,s}(\rho)$, i.e.,
\[
G^k_{q,s}(\rho)
=
\sum_iq_iG^k_{q,s}(|\psi^{(i)}\rangle).
\]

For each pure state $|\psi^{(i)}\rangle$, suppose that the LOCC
operation produces the states
$|\omega^{(il)}\rangle$ with probabilities
\[
p_{il}
=
\operatorname{Tr}
(M_l|\psi^{(i)}\rangle
\langle\psi^{(i)}|M_l^\dagger).
\]
According to the pure-state strong monotonicity proved above, we have
\[
G^k_{q,s}(|\psi^{(i)}\rangle)
\geq
\sum_l p_{il}
G^k_{q,s}(|\omega^{(il)}\rangle).
\]

Therefore,
\begin{equation}\label{iva5-3}
\begin{aligned}
G^k_{q,s}(\rho)
=&
\sum_iq_iG^k_{q,s}(|\psi^{(i)}\rangle)
\\
\geq&
\sum_{i,l}q_ip_{il}
G^k_{q,s}(|\omega^{(il)}\rangle).
\end{aligned}
\end{equation}

The ensemble
\[
\{q_ip_{il},|\omega^{(il)}\rangle\}
\]
is a pure-state decomposition of
\[
\Lambda_{\rm LOCC}(\rho)
=
\sum_{i,l}q_ip_{il}
|\omega^{(il)}\rangle
\langle\omega^{(il)}|.
\]
According to the definition of the convex roof, the minimum average
entanglement over all pure-state decompositions is no larger than the
average entanglement of this particular decomposition. Hence,
\[
\sum_{i,l}q_ip_{il}
G^k_{q,s}(|\omega^{(il)}\rangle)
\geq
G^k_{q,s}
(\Lambda_{\rm LOCC}(\rho)).
\]

Combining the above inequalities, we obtain
\begin{equation}\label{iva5}
G^k_{q,s}(\rho)
\geq
G^k_{q,s}
(\Lambda_{\rm LOCC}(\rho)).
\end{equation}

For each outcome state $\sigma^{(l)}$, let
$\{r_{il},|\phi^{(il)}\rangle\}$ be an optimal decomposition.
Applying the pure-state strong monotonicity to each component gives the strong monotonicity relation
\[
G^k_{q,s}(\rho)
\geq
\sum_l p_l
G^k_{q,s}(\sigma^{(l)}),
\]
where
\[
\sigma^{(l)}
=
\frac{M_l\rho M_l^\dagger}{p_l}.
\]
Therefore, strong monotonicity under LOCC operations is established. Since strong monotonicity implies monotonicity under deterministic
LOCC operations, item (iv) follows immediately.

\section{ The proof of inequalities (\ref{r3-1}) and (\ref{r3-2}) }

Suppose the quantum state $\sigma$ has the spectral decomposition $\sigma=\sum\limits_i\lambda_i|\phi^{(i)}\rangle\langle\phi^{(i)}|$. Then we can define the function $$f(s):=S_{q,s}(\sigma)=\dfrac{A^s-1}{(1-q)s},$$
where $A=\operatorname{Tr}(\sigma^q)=\sum\limits_i\lambda_i^q$ with $q>0$ and $q\neq1$. Since $\lambda_i$ are  eigenvalues of the quantum state $\sigma$,  it follows that $A>0$, and $A=1$ if and only if $\sigma$ is a pure state. We now consider two cases: $A=1$ and $A\neq1$.

Case 1: When $A=1$, $f(s)=0$, and hence $f(s)=0$ is a constant function.

Case 2: When $A\neq1$, let $g(s)=\dfrac{A^s-1}{s},$ then one has
$$g'(s)=\dfrac{sA^s\ln A-A^s+1}{s^2}.$$
Let $r(s)=sA^s\ln A-A^s+1$. Since $\lim_{s\rightarrow0^+}r(s)=0$ and $r'(s)>0$,
we have $r(s)>0$ for $s>0$.
Hence, $g'(s)>0$ for any $s>0$, i.e., $g(s)$ is monotonically increasing in $s$ for $s>0$.
Furthermore, we obtain that for $s>0$, $f(s)$ is monotonically increasing in $s$ when $0<q<1$, and monotonically decreasing in
$s$ when $q>1$.

Combining the above two cases, we obtain that
$$S_{q,s_1}(\sigma)\leq S_{q,s_2}(\sigma)$$
when $0<s_1\leq s_2$ with $0<q<1$, and
$$S_{q,s_1}(\sigma)\geq S_{q,s_2}(\sigma)$$
when $0<s_1\leq s_2$ with $q>1$.
For the limiting case $s=0$, the unified entropy reduces to the R\'enyi entropy,
\[
S_{q,0}(\rho)=R_q(\rho).
\]
The monotonicity with respect to $q$ then follows from the known ordering property of R\'enyi entropies.

\section{Proof of Inequality (\ref{r3-3})}

Let $\Lambda$ be an arbitrary non-empty proper subset of $\{1,2,\ldots,N\}$.
Denote by $\psi_{\Lambda}$ and $\widetilde{\psi}_{\Lambda}$ the vectors consisting of the eigenvalues of the reduced density matrices $\rho_{\Lambda}$ and $\widetilde{\rho}_{\Lambda}$, respectively, where $\rho_{\Lambda}$ and $\widetilde{\rho}_{\Lambda}$ are obtained from $|\psi\rangle$ and $|\widetilde{\psi}\rangle$ on the subsystem $\bigotimes_{i\in\Lambda}H_i$.

To prove inequality (\ref{r3-3}), it is sufficient to establish that
\begin{equation}\label{r3-3-0}
S_{q,s}(\rho_{\Lambda})\geq S_{q,s}(\widetilde{\rho}_{\Lambda})
\end{equation}
for $q>0$, $q\neq 1$, $s>0$, under the assumption
\[\boldsymbol{\lambda}(\rho_\Lambda)
\prec
\boldsymbol{\lambda}(\widetilde{\rho}_\Lambda)\]

Suppose that
\[
\psi_{\Lambda}=(\lambda_1,\lambda_2,\ldots,\lambda_{d_\Lambda}),
\qquad
\widetilde{\psi}_{\Lambda}=(\widetilde{\lambda}_1,\widetilde{\lambda}_2,\ldots,
\widetilde{\lambda}_{d_\Lambda}),
\]
where $\{\lambda_i\}$ and $\{\widetilde{\lambda}_i\}$ are the eigenvalues of $\rho_{\Lambda}$ and $\widetilde{\rho}_{\Lambda}$, respectively. Then
\[
\operatorname{Tr}(\rho_{\Lambda}^{q})
=\sum_{i=1}^{d_\Lambda}\lambda_i^q,
\qquad
\operatorname{Tr}(\widetilde{\rho}_{\Lambda}^{q})
=\sum_{i=1}^{d_\Lambda}\widetilde{\lambda}_i^q .
\]

Since
\[
\boldsymbol{\lambda}(\rho_{\Lambda})
\prec
\boldsymbol{\lambda}(\widetilde{\rho}_{\Lambda}),
\]
Karamata's inequality implies that
\begin{equation}
\operatorname{Tr}(\rho_{\Lambda}^{q})
\leq
\operatorname{Tr}(\widetilde{\rho}_{\Lambda}^{q}),
\qquad q>1,
\end{equation}
because the function $x^q$ is convex for $q>1$. Similarly,
\begin{equation}
\operatorname{Tr}(\rho_{\Lambda}^{q})
\geq
\operatorname{Tr}(\widetilde{\rho}_{\Lambda}^{q}),
\qquad 0<q<1,
\end{equation}
since $x^q$ is concave on $(0,\infty)$ for $0<q<1$.

Recall that the unified $(q,s)$-entropy is given by
\[
S_{q,s}(\rho)
=
\frac{(\operatorname{Tr}\rho^q)^s-1}{(1-q)s}.
\]

For $q>1$, we have
\[
\operatorname{Tr}(\rho_{\Lambda}^{q})
\leq
\operatorname{Tr}(\widetilde{\rho}_{\Lambda}^{q}).
\]
Moreover, for \(s\neq0\), define
\[
f(x)=\frac{x^s-1}{(1-q)s}.
\]
Its derivative is
\[
f'(x)=\frac{x^{s-1}}{1-q}.
\]

Moreover, for $q>1$,
\[
\operatorname{Tr}(\rho_\Lambda^q)
\leq
\operatorname{Tr}(\widetilde{\rho}_\Lambda^q),
\]
and since $f(x)$ is decreasing,
the desired inequality follows.
we obtain
\[
S_{q,s}(\rho_{\Lambda})
=
f\bigl(\operatorname{Tr}\rho_{\Lambda}^{q}\bigr)
\geq
f\bigl(\operatorname{Tr}\widetilde{\rho}_{\Lambda}^{q}\bigr)
=
S_{q,s}(\widetilde{\rho}_{\Lambda}).
\]

For \(0<q<1\), \(f'(x)>0\), and thus \(f(x)\) is monotonically increasing. Therefore, the same trace inequality yields
\[
S_{q,s}(\rho_{\Lambda})
=
f\bigl(\operatorname{Tr}\rho_{\Lambda}^{q}\bigr)
\geq
f\bigl(\operatorname{Tr}\widetilde{\rho}_{\Lambda}^{q}\bigr)
=
S_{q,s}(\widetilde{\rho}_{\Lambda}).
\]

Consequently, inequality (\ref{r3-3-0}) holds for all \(q>0\), \(q\neq1\), and \(s>0\), completing the proof of inequality (\ref{r3-3}).

\end{document}